\documentclass[aps,twocolumn,showpacs,preprintnumbers,nofootinbib,prd,superscriptaddress,10pt]{revtex4-1}

\usepackage[utf8]{inputenc}

\usepackage{graphicx,amssymb,amsmath,amsthm,amsfonts}
\usepackage[usenames]{color}

\usepackage{notes2bib}

\usepackage{aas_macros}
\usepackage{bm}
\usepackage{dcolumn}
\usepackage{latexsym}
\usepackage{rotating}
\usepackage{longtable}

\usepackage{enumerate}
\usepackage{tensor,multirow}
\usepackage{url}
\usepackage[linktocpage]{hyperref}

\def\be{\begin{equation}}
\def\ee{\end{equation}}
\def\beq{\begin{eqnarray}}
\def\eeq{\end{eqnarray}}
\begin{document}

\title{Eccentricity evolution of compact binaries\\ and applications to gravitational-wave physics}

\author{Vitor Cardoso}
\affiliation{CENTRA, Departamento de F\'{\i}sica, Instituto Superior T\'ecnico -- IST, Universidade de Lisboa -- UL,
Avenida Rovisco Pais 1, 1049 Lisboa, Portugal}

\author{Caio F. B. Macedo}
\affiliation{Faculdade de Física, Universidade Federal do Par\'a, Salin\'opolis, Par\'a, 68721-000 Brazil}

\author{Rodrigo Vicente}
\affiliation{CENTRA, Departamento de F\'{\i}sica, Instituto Superior T\'ecnico -- IST, Universidade de Lisboa -- UL,
Avenida Rovisco Pais 1, 1049 Lisboa, Portugal}
\begin{abstract} 
Searches for gravitational waves from compact binaries focus mostly on quasi-circular motion, with the rationale that wave emission circularizes the orbit.
Here, we study the generality of this result, when astrophysical environments (e.g., accretion disks) or other fundamental interactions are taken into account.
We are motivated by possible electromagnetic counterparts to binary black hole coalescences and orbits, but also by the possible use of eccentricity as a smoking-gun for new physics.
We find that: i) backreaction from radiative mechanisms, including scalars, vectors and gravitational waves circularize the orbital motion. 
ii) by contrast, environmental effects such as accretion and dynamical friction increase the eccentricity of binaries.
Thus, it is the competition between radiative mechanisms and environmental effects that dictates the eccentricity evolution.
We study this competition within an adiabatic approach, including gravitational radiation and dynamical friction forces. We show that that there is a critical semi-major axis below which gravitational radiation dominates the motion and the eccentricity of the system decreases. However, the eccentricity inherited from the environment-dominated stage can be substantial, and in particular can affect LISA sources. We provide examples for GW190521-like sources.
\end{abstract}

%%%%
%%%%
\maketitle

%%%%%%%%%%%%%%%%%%%%%%%%%%%%%%
\section{Introduction}
%%%%%%%%%%%%%%%%%%%%%%%%%%%%%%
Merging black hole binaries (BHBs) are now ``visible'', thanks to gravitational-wave (GW) astronomy~\cite{Abbott:2016blz,Barack:2018yly}.
A good modeling of the dynamics of such compact binaries is important to increase our ability to actually see them,
to infer the properties of the merging objects and to impose constraints on the underlying gravitational theory, or other fundamental interactions~\cite{Barack:2018yly}.

It has long been known that orbits which are initially eccentric will quickly circularize on relatively short timescales~\cite{Peters:1964zz,Krolak:1987ofj,Key:2010tc}.
This is true {\it in vacuum}, and thought to describe well stellar mass BHBs, which form substantially prior to merger and evolve mostly only via GW emission.
However, a re-appreciation of eccentricity evolution is required for different reasons. To begin with, the formation of supermassive BHBs is poorly understood. Some of the mechanisms that contribute to such binaries forming and merging actually may also impart a substantial eccentricity,
specially in their initial stages~\cite{Barack:2018yly}. In addition, observations are progressively indicating that large eccentricities may not be rare. One known supermassive BHB (OJ287) was reported to have eccentricity $e\sim 0.65$, while evolving around the disk of the massive component~\cite{Laine:2020dnr}. Such observations were made in the electromagnetic spectrum, but there are indications that some of the GW events, such as GW190521~\cite{Abbott:2020tfl,Abbott:2020mjq} could also originate from eccentric orbits~\cite{Gayathri:2020coq,CalderonBustillo:2020odh}. It is interesting to note that this same event may have an associated electromagnetic counterpart, product of a nontrivial surrounding environment~\cite{Graham:2020gwr}.
A nontrivial environment leads to large center-of-mass drift velocities~\cite{Cardoso:2020lxx} and may lead to large eccentricities during evolution. Even in vacuum, spin-spin couplings at the second post-Newtonian order may induce a nontrivial eccentricity evolution~\cite{Gergely:1998sr,Klein:2010ti,Klein:2018ybm,Phukon:2019gfh}.

The understanding of eccentricity evolution is also important to constrain the presence of new fields. Under the assumption of circular motion, it has been shown that GW observations can impose severe limits on the dipolar moment and charge of the inspiralling objects~\cite{Barausse:2016eii,Cardoso:2016olt}. When the binary components are charged under new fields, emission
in such channels dominates of GW emission at sufficiently low frequencies; hence the assumption that circular remains circular (\textit{i.e.} that radiative processes conspire to circularize the orbit) must be
proved. The purpose of this work is precisely to address the issues above.~\footnote{Throughout this work we use units $c=1$, but we shall write $c$ explicitly in some cases to facilitate the discussion.}

%%%%%%%%%%%%%%%%%%%%%%%%%%%%%%%%%%%%%%%%%%%%%%%%%%%%%%%%%%%%%%
\section{Evolution driven by fundamental fields\label{sec:ff}}
%%%%%%%%%%%%%%%%%%%%%%%%%%%%%%%%%%%%%%%%%%%%%%%%%%%%%%%%%%%%%%
The problem of eccentricity and orbital radius evolution is tightly connected to the ratio of energy to angular momentum loss during the binary evolution.
Take a compact binary of two
objects of mass $m_1, m_2$, and define the total mass and mass ratio
\be
M\equiv m_1+m_2\,,\qquad q=\frac{m_2}{m_1}\,.
\ee
For binaries dominated by the gravitational interaction, the (Newtonian) orbital frequency $\omega_0$ satisfies Kepler's law
\be
\omega_0=\sqrt{\frac{GM}{a^3}}\,,
\ee
where $a$ is the orbital semi-major axis.
In this case, the conserved energy and angular momentum on Keplerian motion are
\beq 
E&=&-\frac{Gm_1m_2}{2a}\,,\label{Kepler_Energy}\\
L^2&=&\frac{Gm_1^2m_2^2a(1-e^2)}{M}\,, \label{Kepler_AngMom}
\eeq
where $e$ is the eccentricity. 

Suppose now that the only decay channel available for the binary evolution is a 
massless field of frequency $\omega$ and azimuthal dependence $e^{im\phi}$. This could be a GW, but could include also a scalar or even a vector field. In this circumstance, then the emitted angular momentum and energy satisfy~\cite{Brito:2015oca}
\be
\frac{\dot{L}^{\rm rad}}{\dot{E}^{\rm rad}}=\frac{m}{\omega}=\frac{1}{\omega_0}\,.\label{special_massless} 
\ee
How do the eccentricity and semi-major axis of the binary
evolve? Energy and angular momentum balance yield
\be
\dot{E}=-\dot{E}^{\rm rad}\leq 0 \,, \quad \dot{L}=-\dot{L}^{\rm rad}\,,
\ee
so we find
\beq
\dot{a}&=&-\frac{2a^2\dot{E}^{\rm rad}}{Gm_1m_2} \leq 0\,,\\
\dot{e}&=&\sqrt{\frac{M}{Ga}}\frac{\sqrt{1-e^2}}{e}\frac{\dot{E}^{\rm rad}}{m_1m_2}\left(\frac{\dot{L}^{\rm rad}}{\dot{E}^{\rm rad}}-\frac{\sqrt{1-e^2}}{\omega_0}\right)\,.
\eeq
We see immediately that, if~$\dot{L}^{\rm rad}/\dot{E}^{\rm rad}$ have eccentricity-dependence starting at order higher than $e^2$, then circular orbits are unstable (\textit{i.e.} $\dot{e}\geq0$ for~$e\sim 0$) on account
of condition \eqref{special_massless}. In case of~$\dot{L}^{\rm rad}/\dot{E}^{\rm rad}$ having eccentricity-dependence starting at order~$e^2$, circular orbits will also be unstable if the coefficient multiplying~$e^2$ is larger than~$-\tfrac{1}{2 \omega_0}$.
%This simple calculation also seems to show that no circular orbits, or small eccentricity orbits are possible when such condition is violated, and this serves as a motivation to study massive fields and general accretion or drag scenarios.

We therefore start our analysis by asking how does the emission of fundamental massless fields affect eccentricity evolution.
%%%%%%%%%%%%%%%%%%%%%%%%%%%%%%%%%%%%%%%%%%%%%%%%
\subsection{Eccentricity evolution in vacuum}
%%%%%%%%%%%%%%%%%%%%%%%%%%%%%%%%%%%%%%%%%%%%%%%%
Let's first assume that our system is in vacuum, isolated from all other sources in the universe.
In this case, the evolution is driven solely by GW emission.
Eccentricity in vacuum GR can be calculated in a two-step procedure.
Take a binary of pointlike objects of mass $m_1, m_2$. To lowest post-Newtonian order,
their motion is elliptical, of semi-major axis $a$ and eccentricity $e$.
Their binding energy $E$ and angular momentum $L$ are simply described by Eqs.~\eqref{Kepler_Energy}-\eqref{Kepler_AngMom}.
Now, when relativistic effects are included, the system radiates energy and angular momentum, via GWs, at a rate
\beq
\langle\dot{E}\rangle&=&-\frac{32}{5}\frac{G^4m_1^2m_2^2M}{a^5(1-e^2)^{7/2}}\left(1+\frac{73}{24}e^2+\frac{37}{96}e^4\right)\,,\\
\langle\dot{L}\rangle&=&-\frac{32}{5}\frac{G^{7/2}m_1^2m_2^2M^{1/2}}{a^{7/2}(1-e^2)^{2}}\left(1+\frac{7}{8}e^2\right)\,.
\eeq
%
%We see that the energy and angular momentum fluxes have eccentricity dependence which when Taylor expanded begins at order $e^2$, and from the previous discussion we expect circular orbits to be stable.
Assuming a slow, adiabatic evolution, one can now follow Peters~\cite{Peters:1964zz} and compute the major axis and eccentricity evolution. For small eccentricity, one finds
\beq
\langle\dot{a}\rangle&=&-\frac{64 G^3}{5}\frac{m_1m_2M}{a^3}<0\,,\\
\langle\dot{e}\rangle&=&-\frac{304G^3}{15}\frac{m_1m_2M}{a^4}\,e \leq 0\,.
\eeq
In other words, the major axis decreases with time due to energy loss in GWs. So does the eccentricity, thus orbits tend to become circular on long timescales.
Note, however, that eccentricity evolution is very sensitive, in particular, it hardly evolves for quasi-circular orbits.
One is thus forced to consider what happens when other physics sets in.

%%%%%%%%%%%%%%%%%%%%%%%%%%%%%%%%%%%%%%%%%%%%%%%%%%%%%%%%%%%%%%%%%%%%%
\subsection{Evolution in the presence of scalar and vector radiation}
%%%%%%%%%%%%%%%%%%%%%%%%%%%%%%%%%%%%%%%%%%%%%%%%%%%%%%%%%%%%%%%%%%%%%
Consider, then, binary components carrying some additional charge. The simplest examples include scalar charge, as is the case in scalar-tensor theories, or 
electromagnetic charge (the theory below also describes some dark matter models with mili-charged components~\cite{Cardoso:2016olt}). We model this via the theory of massless fields
\beq 
\mathcal{S}&=&\int d^4x \sqrt{-g}\bigg[\frac{R}{8 \pi G}-g^{\mu \nu}\Phi_{,\mu} \Phi_ {,\nu}-\frac{1}{2}F^{\mu\nu}F_{\mu\nu}\nonumber\\
&-&\frac{2}{\sqrt{-g}}\sum_{j=1}^{2}(m_j+ 4 \pi q^{0}_j \Phi) \int d\lambda \sqrt{-g_{\mu \nu}\dot{z_j}^\mu \dot{z_j}^\nu}\delta^4(x-z_j)\nonumber\\
&-&\frac{8\pi}{\sqrt{-g}}\sum_{j=1}^2 q^{1}_j A_\alpha \int d\lambda\, \dot{z}_j^\alpha\delta^4(x-z_j)\bigg]\,.\label{Action_total}
\eeq
Here, $\Phi$ is a massless scalar, $A_\mu$ is a massless vector and the Maxwell tensor $F_{\mu\nu}=\nabla_{\mu} A_{\nu}-\nabla_{\nu} A_{\mu}$.
Each of the binary components carries a charge $q^{s}_i$ of the corresponding spin-$s$ field ($s=0,1$ for scalar and vectors, respectively).

The details of the calculation are shown in Appendix \ref{sec:details_emission}. As might be anticipated, in the weak field regime the motion is Keplerian with energy and angular momentum 
\be
E=-\frac{\tilde{G} m_1m_2}{2a}\,,\qquad L^2=\frac{\tilde{G}m_1^2m_2^2a(1-e^2)}{M}\,,\label{eq:energy_angularmomentum_main}
\ee
where the effective Newton's constant is now
\begin{align}
	\tilde{G}\equiv G-4 \pi \frac{q^s_1 q^s_2}{m_1 m_ 2}\,,
\end{align}
where we assume (without loss of generality) that only one further interaction ($s=0$ {\it or} $s=1$) is turned on.

In the Newtonian approximation, radiation propagates in flat space and the Green's function for the problem is well known.
Averaging over an orbit, we find the surprisingly compact expressions for the rate of energy and angular momentum emission
\begin{align}
&\langle\dot{E}^{\rm rad}\rangle=\frac{2\pi (s+1)}{3}\frac{\tilde{G}^2}{a^4}(q^s_1 m_2-q^s_2 m_1)^2 \left(\frac{2+e^2}{(1-e^2)^{\frac{5}{2}}}\right) \label{EnergySavp} \,, \\
&\langle\dot{L}^{\rm rad}\rangle=\frac{4 \pi (s+1)}{3} \frac{\tilde{G}^\frac{3}{2}}{\sqrt{M}a^{\frac{5}{2}}(1-e^2)}(q^s_1 m_2-q^s_2 m_1)^2  \label{AngMomSavp}\,,
\end{align}
resulting in the spin-independent dipolar ratio
\begin{align}
	&\frac{\langle\dot{L}^{\rm rad} \rangle}{\langle\dot{E}^{\rm rad} \rangle} = \frac{\sqrt{1-e^2}}{\omega_0} \left(\frac{1-e^2}{1+\frac{e^2}{2}}\right) \,.\label{LEratioSp}
\end{align}
The flux of scalar energy in the circular orbit limit agrees with that of Refs.~\cite{Cardoso:2011xi,Yunes:2011aa,Cardoso:2019nis}.
Our results for the electromagnetic flux of energy and angular momentum agree with those in Refs.~\cite{Christiansen:2020pnv,Liu:2020cds} (after a proper re-definition of charge).
In the adiabatic approximation the major semi-axis and the eccentricity follow
\begin{align}
	\langle\dot{a}\rangle&=-\frac{2a^2\langle\dot{E}^{\rm rad}\rangle}{\tilde{G} m_1m_2}<0\,,\label{shrinkp}\\
	\langle\dot{e}\rangle&=\sqrt{\frac{M}{ \tilde{G}a}}\frac{\sqrt{1-e^2}}{e}\frac{\langle\dot{E}^{\rm rad}\rangle}{m_1m_2}\left(\frac{\langle\dot{L}^{\rm rad}\rangle}{\langle\dot{E}^{\rm rad}\rangle}-\frac{\sqrt{1-e^2}}{\omega_0}\right) \nonumber \\
	&=-\sqrt{\frac{M}{ \tilde{G}a}}\left(\frac{1-e^2}{e\, \omega_0}\right)\frac{\langle\dot{E}^{\rm rad}\rangle}{m_1m_2}\left(\frac{3 e^2}{2+e^2}\right)\leq 0\,.\label{circularp}
\end{align}
Thus, the emission of massless radiation by a binary causes the major semi-axis and the eccentricity to decrease in time: the orbit shrinks and circularizes. 
Although we will not explore the subject further, it is important to realize that electromagnetic fields couple strongly to plasmas. Thus, when applied to the Maxwell sector, the previous results should be taken with care~\cite{Cardoso:2020nst}.

%%%%%%%%%%%%%%%%%%%%%%%%%%%%%%%%%%%%%%%%%%%%%%%%%%%%%%%%%%%%%%%%%%%%%%%%%%%%%%%%%%%%%%%%%%%%%%%%%%%%
\section{Eccentricity evolution in constant-density environments: accretion and dynamical friction}
%%%%%%%%%%%%%%%%%%%%%%%%%%%%%%%%%%%%%%%%%%%%%%%%%%%%%%%%%%%%%%%%%%%%%%%%%%%%%%%%%%%%%%%%%%%%%%%%%%%%

The presence of surrounding dust or plasma affects the above picture in different ways. Binaries, such as the event GW190521~\cite{Abbott:2020tfl,Abbott:2020mjq}, may in fact evolve within accretion disks, where the density of the surrounding environment may play an important role.
The presence of matter surrounding a BHB will cause accretion to occur~\cite{Bondi:1944jm,Macedo:2013qea,Edgar:2004mk}. A second mechanism at play is dynamical friction (DF), whereby the moving BHs get dragged down by the surrounding matter~\cite{Chandrasekhar:1943ys,Ostriker:1998fa,Annulli:2020lyc,Macedo:2013qea}. 

Consider first accretion. We assume that the surrounding medium has constant density. This implies in particular that there is a supply mechanism
that keeps the density constant even as the binary sweeps through and accretes some of the particles.
We neglect here the gravitational potential generated by the accretion disk or surrounding matter; this approximation is expected to be extremely good for BHBs close to merger. We focus on Bondi-Hoyle accretion~\cite{Edgar:2004mk}. The mass flux at the horizon is
\beq
\dot{m}_i=4\pi G^2\rho \frac{m_i^2}{(v_i^2+c_s^2)^{3/2}}\,,
%
%\dot{m}_i^{\rm collisionless}&=&\frac{4\pi G^2}{c^2}\rho \frac{m_i^2}{v_i}\,,
\eeq
when the binary components are BHs. These are Newtonian formulas, expected to be valid up to factors of order 1 when the binary is non-compact. Here, $v_i$ is the relative velocity between BH ``$i$'' and the environment, and $c_s$ is the sound speed in the medium. We will always consider regimes for which $v_i\gg c_s$. Numerical studies indicate that the above description is solid, even in the presence of wake instabilities~\cite{Edgar:2004mk}.

Binaries in a medium are also subject to the gravitational force due to the wakes generated by the moving bodies, as we mentioned. This DF depends on the characteristics of the fluid and on the moving bodies. In summary, DF can usually be represented by a external force of the type
\beq
\mathbf{F}_{{\rm d},i}=-G^2m_i^2\rho I_{\rm d}(v_i)\dot{\mathbf{r}}_i\,,\label{eq:accretionm}
\eeq 
where the form of the function $I_{\rm d}$ depends on the specifics of the DF model at hand.
We consider the dynamical friction in a fluid (collisional) medium in the supersonic regime ($v_i\gg c_s$), for which~\cite{Dokuchaev:1964,Ruderman:1971,Rephaeli:1980,Ostriker:1998fa}~\footnote{This expression assumes linear motion in an extended medium. The fact that the binary components do not follow a linear motion and are inside a (possibly thin) disk introduces some modifications to the DF, which we neglect here for simplicity. For a more careful analysis of the DF in these type of systems, we direct the reader to,~\textit{e.g.} Ref.~\cite{Antoni:2019pgq,Vicente:2019ilr}.}
\be
I_{\rm d}(v_i)= \frac{4\pi\lambda}{v_i^3}\,,
\ee
where $\lambda$ is the Coulomb logarithm. It is easy to see that, for large velocities, the Chandrasekhar formula for collisionless media~\cite{Chandrasekhar:1943ys} reduces to the last expression. 
We adopt $\lambda\sim20$, unless stated otherwise, but note that changing $\lambda$ is equivalent to re-normalizing the density in the DF expression. As we show below, even a factor 10 variation in this parameter has only a mild effect on the overall evolution of the system.
%We focus on the Chandrasekhar model to describe dynamical friction~\cite{Chandrasekhar:1943ys}, for which
%\be
%I_{\rm d}(v_i)= \frac{4\pi\lambda}{v_i^3}\left[{\rm erf}(v_i/(\sqrt{2}\sigma))-\frac{2v_i}{\sqrt{2\pi}\sigma}e^{-v_i/(2\sigma^2)}\right],
%\ee
%where $\lambda$ is the Coulomb logarithm, and $\sigma$ the dispersion of the matter Maxwellian velocity distribution. For similar reasons as above, we also restrict attention to the high-speed limit, namely $v_i\gg\sigma$. We also adopt $\lambda\sim20$, unless stated otherwise, but note that changing $\lambda$ is equivalent to renormalizing the density. As we show below, even a factor 10 variation in this parameter has only a mild effect on the overall evolution of the system.

Taking then a binary evolving under the influence of accretion and DF, the equations of motion can be written as
\begin{align}
m_i \ddot{\mathbf{r}}_i+\dot{m}_i \dot{\mathbf{r}}_i=\pm\frac{G m_1m_2}{r^3}\mathbf{r}+\mathbf{F}_{{\rm d},i}\,,\label{eq:r1r2}
\end{align}
where $\mathbf{r}=\mathbf{r}_2-\mathbf{r}_1$ is the orbital separation vector of the binary. Introducing the center of mass of the binary
\begin{equation}
	\mathbf{R}=\frac{m_1\mathbf{r}_1+m_2\mathbf{r}_2}{m_1+m_2}\,,
\end{equation}
we can write a system of equations describing the vectors $\mathbf{r}$ and $\mathbf{R}$, namely
\begin{align}
	\ddot{\mathbf{r}}&=f_{1}\dot{\mathbf{r}}+f_{2}\dot{\mathbf{R}}+f_{3}{\mathbf{r}}\,,\label{eq:eqr}\\
	\ddot{\mathbf{R}}&=f_{4}\dot{\mathbf{r}}+f_{5}\dot{\mathbf{R}}+f_{6}{\mathbf{r}}\,,\label{eq:eqR}
\end{align}
where the functions $f_i$ are given by
\begin{widetext}
\begin{align}
f_1&=-\frac{G^2 M q \rho  (I_{\text{a1}}+I_{\text{a2}}+I_{\text{d1}}+I_{\text{d2}})}{(q+1)^2}\,,\\
f_2&=\frac{G^2 M \rho  [I_{\text{a1}}+I_{\text{d1}}-q (I_{\text{a2}}+I_{\text{d2}})]}{q+1}\,,\\
f_3&=G M \left\{\frac{G^3 M q \rho ^2 (I_{\text{a1}}-q I_{\text{a2}}) [I_{\text{a1}}+I_{\text{d1}}-q
		(I_{\text{a2}}+I_{\text{d2}})]}{(q+1)^4}-\frac{1}{r^3}\right\}\,,\\
f_4&=\frac{G^2 M q \rho  [q (I_{\text{a2}}-I_{\text{d2}})-I_{\text{a1}}+I_{\text{d1}}]}{(q+1)^3}\,,\\
f_5&=-\frac{G^2 M \rho  \left[q^2 (I_{\text{a2}}+I_{\text{d2}})+I_{\text{a1}}+I_{\text{d1}}\right]}{(q+1)^2}\,,\\
f_6&=-\frac{G^4 M^2 q \rho ^2 (I_{\text{a1}}-q I_{\text{a2}}) \left[q^2 (I_{\text{a2}}+I_{\text{d2}})+2 q
		(I_{\text{a1}}+I_{\text{a2}})+I_{\text{a1}}+I_{\text{d1}}\right]}{(q+1)^5}\,.
\end{align}
\end{widetext}
Here, we defined
\begin{equation}
I_{ai}=\frac{4\pi}{(v_i^2+c_s^2)^{3/2}},~I_{di}=I_d(v_i)\,.
\end{equation}
Note that due to accretion, both the mass-ratio and the total mass evolve in time. We can compute their evolution via Eq.~\eqref{eq:accretionm}, obtaining
\begin{align}
\dot{q}&=\frac{G^2 M q \rho  (q I_{\text{a2}}-I_{\text{a1}})}{q+1},\label{eq:q}\\
\dot{M}&=\frac{G^2 M^2 \rho  \left(q^2 I_{\text{a2}}+I_{\text{a1}}\right)}{(q+1)^2}.\label{eq:M}
\end{align}

To investigate the evolution of the system, equations \eqref{eq:eqr}, \eqref{eq:eqR}, \eqref{eq:q}, and \eqref{eq:M} must be solved together. Note that the equations for the center of mass vector predict a boost, as can be seen in~\cite{Cardoso:2020lxx}. To analyze the eccentricity evolution, however, we have to focus into $\mathbf{r}$ instead. Before going into the full regime, it is instructive to focus on some particular cases.

\subsection{Equal-mass binaries}
%%%%%%%%%%%%%%%%%%%%%%%%%%%%%%%
For equal mass ratio binaries, $q=1$ during the whole evolution, due to symmetry [c.f. Eq.~\eqref{eq:q}]\footnote{We note that we are considering a homogeneous medium. Density lumps in the medium can introduce asymmetries that can affect the outcome of the motion.}. In this case, the center of mass remains at rest (or constant velocity) and the equations simplify considerably. Considering $\mathbf{R}=0$, we have
\begin{equation}
\ddot{\mathbf{r}}=-\frac{G^2M\rho}{2}(I_a+I_v)\dot{\mathbf{r}} - \frac{G M}{r^3}\mathbf{r}\,,\label{eq:r_eq}
\end{equation}
where we dropped the particle label index because drag and accretion forces are the same for both particles. Additionally, the total mass of the particles also evolves because of accretion. The total mass evolution is given by
\begin{equation}
\dot{M}=\frac{G^2M^2\rho I_a}{2}\,.\label{eq:M_eq}
\end{equation}

To track the eccentricity of the system, it is useful to describe the evolution of the total mechanical energy and the angular moment of the reduced mass. The evolution of the mechanical energy can be found by analyzing the power extracted by the external force. We have that the energy per unit of reduced mass is determined by
\begin{equation}
\dot{\varepsilon}=-\frac{G^2M\rho(I_a+I_v)}{2}\dot{\mathbf{r}}\cdot\dot{\mathbf{r}} =-\frac{G^2 M\rho k}{2v}\,,\label{eq:dote}
\end{equation}
where $v=|\dot{\mathbf{r}}|$, and we considered $(I_a+I_v)\approx k/v^3$, which is valid even for collisional DF in the limit $v/c_s\gg1$.
\footnote{
For the model adopted here, considering only DF, we have $k=32\pi \lambda$ (note that $v_i=v/2$ for symmetric binaries).
} The evolution of the angular momentum per reduced mass ($|\mathbf{r}\times\dot{\mathbf{r}}|$) follows from the differential Eq.~\eqref{eq:r_eq},
\begin{equation}
\dot{h}=-\frac{G^2 M\rho k}{2v^3}h\,.\label{eq:doth}
\end{equation}
Finally, the eccentricity can be found by tracking
\begin{equation}
e=\sqrt{1+2\frac{\varepsilon h^2}{G^2M^2}}.
\label{eq:ecc}
\end{equation}

%
%As noted previously, in many astrophysical scenarios $v\gg c_s$. For simplicity, we can use $I_a+I_v=k_1/v^3$ $I_a=k_2/v^3$. When dynamical friction is subdominant, we have $k_1\approx k_2$. By using the angular momentum per unit of reduced mass as $h=r^2\dot{\varphi}$, and $u\equiv 1/r$, one can find the following equations
%\begin{align}
%	u''&=\frac{G M}{h^2}-u,\\
%	h'&=-\frac{4 G^2 k_1 \rho  M}{h^3 u^2 \left(u'^2+u^2\right)^{3/2}},\\
%	M'&=\frac{4 G^2 k_2 \rho  M^2}{h^4 u^2 \left(u'^2+u^2\right)^{3/2}},	
%\end{align}
%where the prime denotes derivative with respect to $\varphi$ ($'\equiv d/d\varphi$). In this regime, the equations describing the volution of $u$ is very similar to the case without dissipative forces, with the difference that $M$ and $h$ now also evolve with $\varphi$. By using the above equations, we obtain the following relation between $h$ and $M$
%\begin{equation}
%	\frac{h'}{M'}=-\frac{k_1 h}{k_2 M},
%\end{equation}
%which can be readily integrated, obtaining
%\begin{equation}
%	h=h_0\left(\frac{M}{M_0}\right)^{-k_1/k_2},
%\end{equation}
%implying that as the mass increases the angular momentum decreases. This relation can be used to reduce the system to only two differential equations for $(u,h)$. Neglecting dynamical friction, $k_1=k_2$, we have
%\begin{align}
%	u''&=\frac{G h_0 M_0}{h^3}-u,\\
%	h'&=\frac{4 G^2 k_1 \rho  h_0M_0}{h^4 u^2 \left(u'^2+u^2\right)^{3/2}}
%\end{align}

%%%%%%%%%%%%%%%%%%%%%%%%%%%%%%%%%%%%%%%%%%%%%%%%%%%%%%%%%%%%%%%%%%%%%%%%%%%%%%%%%%%%%%%
\subsubsection{Averaging the energy and angular momentum evolution for elliptic orbits}
%%%%%%%%%%%%%%%%%%%%%%%%%%%%%%%%%%%%%%%%%%%%%%%%%%%%%%%%%%%%%%%%%%%%%%%%%%%%%%%%%%%%%%%
In a similar fashion to that of Section~\ref{sec:ff} where we dealt with fundamental fields, we can consider Eqs.~\eqref{eq:dote} and \eqref{eq:doth} as ``fluxes" in which the RHS is computed for a fixed orbit. For simplicity, let us consider only DF, i.e. $M$ is constant during the evolution. For an elliptical orbit, using the average defined in Appendix~\ref{sec:details_emission}, we find the energy and angular momentum loss for one complete cycle
\begin{align}
\left<\dot{\varepsilon}\right>&=-\frac{a \left(1-e^2\right)^2 G k \rho  \sqrt{\frac{G M}{a}}}{4 \pi }\int_0^{2\pi}d\varphi\, g_\varepsilon,\label{eq:vare}\\
\left<\dot{h}\right>&=-\frac{a^2(1-e^2)^{7/2}G k \rho}{4\pi}\int_0^{2\pi}d\varphi\, g_h,\label{eq:varh}\\
g_h&=(1+e\cos\varphi)^{-2}(1+e^2+2e\cos\varphi)^{-3/2},\\
g_\varepsilon&=(1+e\cos\varphi)^{-2}(1+e^2+2e\cos\varphi)^{-1/2}.
\end{align}
Finally, we can use the following relations
\begin{equation}
a=-\frac{GM}{2\varepsilon},~~e^2=1-2\frac{\varepsilon \,h^2}{G^2M^2},
\end{equation}
to rewrite Eqs.~\eqref{eq:vare}-\eqref{eq:varh} in terms of $a$ and $e$. For low-eccentricity orbits, we find
\begin{align}
\left<\dot{a}\right>&=-k \rho  \sqrt{\frac{G\,a^5 }{M}}\left(1+\frac{3 e^2}{4}+{\cal O}(e^4)\right),\\
\left<\dot{e}\right>&=\frac{3}{2} k \rho  \sqrt{\frac{G\,a^3}{M}}e\left(1+\frac{3 e^2}{8}+{\cal O}(e^4)\right).
\end{align}
From the above relations, we see that eccentricity \textit{increases} in time under the effect of the dissipative environmental forces. This has been observed in some works considering motion under the influence of drag~\cite{Gair:2010iv,Macedo:2013qea,Cardoso:2020lxx}.

Using the formalism of adiabatic invariants (see \textit{e.g.}~\cite{landau1982mechanics}) one may be led to expect eccentricity to be constant under the adiabatic approximation (which would contradict some of the results discussed here). While eccentricity is a constant at leading order, the semi-major axis does evolve one this time scale, and some conclusions can be drawn for GW binary systems~\cite{DeLuca:2020qqa}. Although eccentricity is indeed an adiabatic invariant at leading order, it does not need to be (and it is not, in general) a constant of motion at next-to-leading order~\cite{1985Salmassi,1993Djukic}. Additionally, under the regime of validity of the adiabatic approximation, it is true that the eccentricity must change over a timescale much larger than, for instance, the semi-major axis (which is not a constant of motion at leading order). We have verified that eccentricity indeed increase by considering, for instance, a system subject to only accretion-driven forces (which is subdominant over DF), with the evolution of $e(a)$ converging for $\rho\to 0$, indicating that indeed eccentricity does change adiabatically.

%%%%%%%%%%%%%%%%%%%%%%%%%%%%%%%%%%%%%%%%%%%%%%%%%%%%%%%%%%%%%%%%%%%%%%
\subsubsection{Dissipative forces, GWs and the eccentricity evolution}
%%%%%%%%%%%%%%%%%%%%%%%%%%%%%%%%%%%%%%%%%%%%%%%%%%%%%%%%%%%%%%%%%%%%%%
%
\begin{figure*}[ht]
	\includegraphics[width=0.5\linewidth]{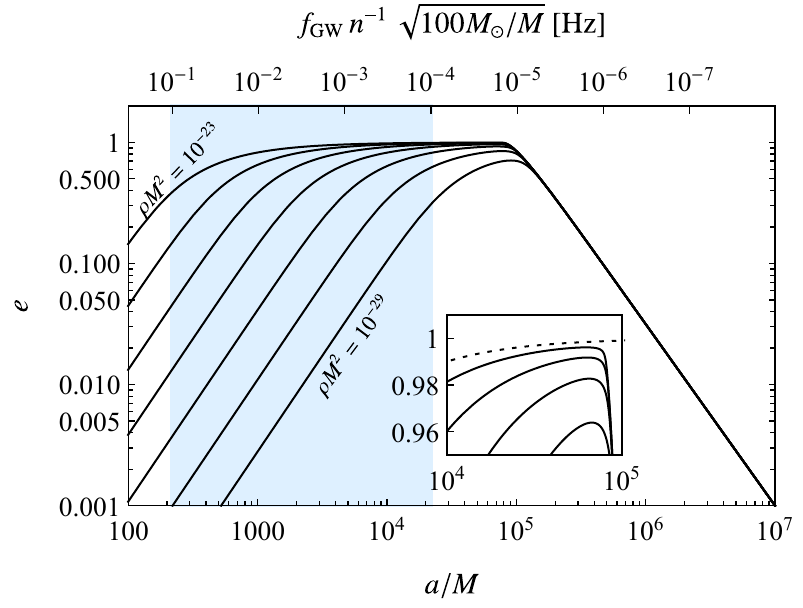}\includegraphics[width=0.5\linewidth]{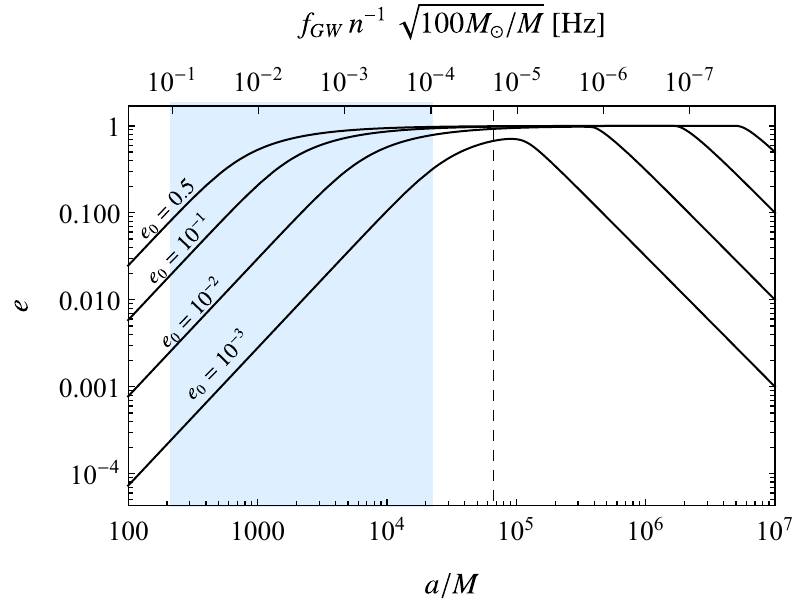}
	\caption{Eccentricity evolution of a binary system, with an initial semi-axis $a/M=10^{7}$. Bottom axis shows the semi-major axis as function of eccentricity, top axis shows the GW frequency. We run the binary up to a distance of $a=100M$. Blue bands indicate LISA's frequency range~\cite{Audley:2017drz}. \textit{Left panel:} We consider a system with an initial eccentricity of $e=10^{-3}$ and different values of the environment density. Dashed line in inset shows threshold values for which periastron is $100M$. \textit{Right panel:} We fix the density to be $\rho M^2=10^{-29}$, changing the initial eccentricity of the system. The vertical line indicates the critical distance, given by Eq.~\eqref{eq:criticala}. }
	\label{fig:ecc_evol}
\end{figure*}
As seen above, dissipative forces such as DF increase the orbital eccentricity of the binary. On the other hand, radiative mechanisms, such as GW emission, act to decrease the orbital eccentricity. We now quantify the combined effect, to understand how binaries behave in astrophysical environments, focusing in the GW channel only. 
We can use the equations for $\left<\dot{a}\right>$ and $\left<\dot{e}\right>$ to compute $da/de$. When only GW emission contributes~\cite{Peters:1964zz,MaggioreBook},
\be
\frac{da}{de}=\frac{12a}{19e}\left(1+\frac{3323}{912}e^2+{\cal O}(e^4)\right) ~~~ (\text{GW-only})\,.
\label{eq:gravonly}
\ee
On the other hand, DF alone produces
\begin{equation}
\frac{da}{de}=-\frac{2a}{3e}\left(1+\frac{3}{8}e^2+{\cal O}(e^3)\right)~~~ (\text{DF-only})\,.
\label{eq:dfecc}
\end{equation}
Curiously, the DF result (expressed in this way) does not depend explicitly on the medium density. At linear order, we can combine the effects of GW emission and DF by simply adding the energy and angular momentum loss, and find, up to terms of order ${\cal O}(e^0)$,
\begin{equation}
\frac{da}{de}=\frac{6 a \left(5 c^5 k \rho  \sqrt{G M a^{11}}+32 G^3 M^4\right)}{e \left(304 G^3 M^4-45 c^5 k \rho  \sqrt{G M a^{11}}\right)}(\text{GW+DF})\,.	\label{eq:gwdf}
\end{equation} 
Interestingly, when the two effects are combined the density of the medium manifests itself. This is because the density balances the contribution from the energy and angular momentum loss. For $\rho=0$, we recover the standard GW case. Clearly, there is a critical value for the distance as function of the medium density in which $da/de$ changes sign. We have
\begin{align}
\frac{a_{\rm c}}{\left( \frac{ 100 G M_\odot}{c^2}\right)}&=3 \times 10^{4}\,k^{-2/11}\left(\frac{M}{100 M_\odot}\right)^{7/11}\left(\frac{\rho_{10}}{\rho}\right)^{2/11},
\label{eq:criticala}
\end{align}
where $\rho_{10}=10^{-10}{\rm g \,cm^{-3}}$.
For $a\lesssim a_c$, GW emission is dominant over DF and the eccentricity decreases. The factor $k^{-2/11} \in [0.1,0.5]$ for most reasonable scenarios
\footnote{Considering $\lambda\in [0.5,2000]$.}.

The critical distance given by Eq.~\eqref{eq:criticala} dictates the balance between environmental forces and GW emission, indicative of whether quasi-circular orbits are indeed expected close to coalescence. However, other factors may be important. One of them is the adiabatic assumption (explored in the Appendix~\ref{sec:adiabatic}, where we show evidence that it does not impact our findings substantially), the other concerns the eccentricity evolution, which depends on the initial conditions and which may lead to extremely small periastron distances.

Figure~\ref{fig:ecc_evol} shows the result of the integration of Eq.~\eqref{eq:gwdf}, including corrections for the DF part up to order ${\cal O}(e^{12})$. We focus on initial semi-major axis of $a(e_0)=10^7M$, for different values of the medium density and the initial eccentricity of the system, but the results hold for other initial distances, observing as the density scales with the separation of the system. Note that
\be
\frac{G^3}{c^6}\rho M^2=1.6\times 10^{-24}\frac{\rho}{\rho_{10}}\left(\frac{M}{100M_{\odot}}\right)^2\,,\label{eq:density_normalization}
%\frac{c^2 r}{G M}&=4.75 \times 10^5\frac{M_\odot}{M}\frac{r}{R_\odot}. \label{eq:conversion_m}
\ee
where we used values typical of event GW190521~\cite{Abbott:2020tfl,Abbott:2020mjq,Graham:2020gwr} as reference values.

It is clear from the figure that the eccentricity increases when the environmental effects dominate, for separations larger than those in Eq.~\eqref{eq:criticala}. In this region $e\propto (a/M)^{-3/2}$, regardless of the medium density and of the initial eccentricity, as predicted by Eq.~\eqref{eq:dfecc}. It is also important to note that, while for small separations GW drives the process with $e\propto (a/M)^{19/12}$, the eccentricity inherited from the environment-dominated phase may be substantial. Thus, the system could still be observed with a considerable eccentricity in a wide range of binary evolution stages. 
Note that $\rho M^2\sim 10^{-22}$ or larger are possible close to the inner edge of thin accretion disks, thus eccentricities larger than $e\sim 0.1$ are expected during a substantial portion of the time-in band for a detector such as LISA.

It is instructive to understand the initial and final stages of the binary evolution analytically. As indicated previously, the GW and medium dominated regions can be estimated by looking into their respective solutions for low eccentricities [i.e., Eqs.~\eqref{eq:gravonly} and~\eqref{eq:dfecc}]. The link between the two regimes can be estimated by analyzing Eq.~\eqref{eq:gwdf}, imposing the initial eccentricities $e_0=e(a_0)$. Let us assume that the motion starts far from the critical distance~\eqref{eq:criticala}. We obtain the following simple expressions for the two regimes
\begin{equation}
	e=\left\{
	\begin{array}{ll}
		e_0\left(\frac{a}{a_0}\right)^{-3/2}, & a\gg a_c,\\
		0.35\,e_0\,\tilde{a}_0^{3/2} \tilde{a}^{19/12}(k \,\tilde{\rho})^{37/66}, & a\ll a_c,
	\end{array}
	\right.
\end{equation}
with $\tilde{a}=a/(G M/c^2)$, and $\tilde{\rho}=G^3M^2\rho/c^{6}$. 	The above solutions are valid mostly for low densities and low initial eccentricities. These expressions can be used to understand all of the peculiarities of Fig.~\ref{fig:ecc_evol}.

For very large eccentricities, it is conceivable that the distance of closest approach would be so small that the components would effectively collide. For the systems we explored, this possibility is not realized. The minimum distance $r_{\rm min}$ obeys
\begin{equation}
r_{\rm min}>100\frac{G M}{c^2},
\end{equation}
which can be translated to maximum eccentricity of $e=1-100(G M/c^2)/a$, represented by the dashed line in the inset of the left panel of Fig.~\ref{fig:ecc_evol}. This indicates that we can expect the objects to pass relatively close to each other without colliding during the evolution, for the density range investigated in the figure. Interestingly, this collision avoidance is only possible due to the GW effect of decreasing the binary eccentricity: If only the medium effects were in play, the objects would collide much sooner and during a highly eccentric motion.

Newtonian circular binaries emit GWs at a frequency $f_{\rm GW}=\omega_0/\pi$. Eccentricity makes the spectrum more complex.
Elliptical orbits will in general generate a spectrum
\be
f_{\rm GW}=n\frac{\omega_0}{2\pi},~{\rm with}~n\geq1.
\ee
Therefore, in general, all harmonics of the orbital frequency contribute to the GW frequency. The dominant frequency, or equivalently the $n=\bar{n}$, depends on the eccentricity of the system. The higher the eccentricity, the higher the value of $\bar{n}$. In other words, high-frequency bursts are emitted at periastron~\cite{Hopper:2017qus}, which means in practice that the source can enter the LISA band much sooner than what seems to be implied by the figure. In Fig.~\ref{fig:ecc_evol} we also show the frequency of the system normalized by the value of $n$. We highlight that the frequencies fall into the LISA band while having a considerable eccentricity.

%%%%%%%%%%%%%%%%%%%%%%%%%%%%%%%%%%%%%%%%%%%%%%%%%%%%%%%%%%%%%%%%
\subsection{Asymmetric binaries and accretion}
%%%%%%%%%%%%%%%%%%%%%%%%%%%%%%%%%%%%%%%%%%%%%%%%%%%%%%%%%%%%%%%%
To implement the simple adiabatic approximation described in the previous sections, we have focused on symmetric binaries and neglected accretion. This approximation enabled us to understand the evolution under the effect of both dynamical friction and GW backreaction. However, asymmetry leads to novel, important effects. It was realized recently that unequal-mass binaries may acquire a large center-of-mass velocity as the evolution proceeds~\cite{Cardoso:2020lxx}. We can also verify here that accretion might not play a central role in the earlier stages of eccentricity gain.

\begin{figure}
	\includegraphics[width=\linewidth]{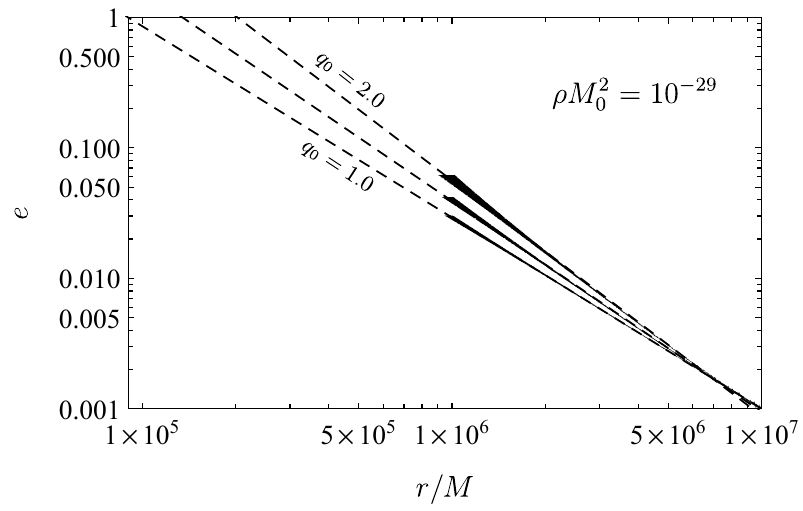}
	\caption{Eccentricity evolution for different initial mass-ratios ($q=1.0,~1.5$ and $2.0$), when accretion is included. The dashed line is an analytical fit that enable us to predict at which distance the system will reach highly eccentric motion.}
	\label{fig:eccenti}
\end{figure}
In order to understand asymmetric binaries and the influence of accretion, we integrate the full system of equations given by Eqs. \eqref{eq:eqr}-\eqref{eq:eqR} and \eqref{eq:q}-\eqref{eq:M}, neglecting possible GW backreaction into the system. This approximation should be valid far from the critical distance \eqref{eq:criticala}, where the environmental effects dominate over GW. We also focus in a regime in which the adiabatic approximation is valid for symmetric binaries in the absence of accretion.

In Fig.~\ref{fig:eccenti} we plot the eccentricity as function of the orbital distance for a medium with density $\rho M^2=10^{-29}$, with initial separation major semi-axis $a_0=10^7M$ and eccentricity $e=0.001$. We verify that the results remain essentially the same for $\rho M^2\in [10^{-28},10^{-30}]$, indicating that we are in the regime in which the adiabatic approximation is valid (see Appendix~\ref{sec:adiabatic}). We also consider initial mass-ratios $q=1,~1.5$, and $2$. For higher mass-ratios eccentricity grows faster as the distance decreases, which is evident by analyzing the slope of the curves in Fig.~\ref{fig:eccenti}. We also display this eccentricity growth by using a fit (dashed lines in Fig.~\ref{fig:eccenti}) to extrapolate the evolution data up to higher eccentricities. This implies that asymmetric binaries will reach highly eccentric motion faster than symmetric ones. 

Accretion has little impact in the evolution of eccentricity, when compared to dynamical friction, for the density range considered in this paper. However, we should highlight that this is model-dependent: To perform the computations, we fix the DF model with $\lambda=20$. In general, in the high-velocity limit, the ratio between the DF force and accretion force is $\lambda$ and, as such, $\lambda=20$ indicates a medium in which dynamical friction generally dominates over accretion. Additionally, because $\lambda$ appears combined with the medium density in the DF force, it also influences the density scales in which the orbits evolve adiabatically.

%%%%%%%%%%%%%%%%%%%%%%%%%%%%%%%%%%%%%%%%%%%%%%%%%%%%%%%%%%%%%%%%%%%%%%%%%%%%%
\section{Discussion}
%%%%%%%%%%%%%%%%%%%%%%%%%%%%%%%%%%%%%%%%%%%%%%%%%%%%%%%%%%%%%%%%%%%%%%%%%%%%%
We studied the evolution of eccentricity of compact binaries, evolving via emission of massless fields and of environmental
accretion and gravitational drag. We proved that the emission of massless scalars, vectors of tensors circularizes the orbits.
In particular, the critical distance at which the orbits start to circularize is larger when additional scalar or vector charges are considered. The integration of 
Eqs.~\eqref{shrinkp}-\eqref{circularp} shows that
\be
a/M=c\frac{e^{4/3}}{1-e^2}\,,
\ee
with $c$ a constant, for scalar or vector-driven binaries. Compare this against the gravitational-driven result, $a/M\sim ce^{12/19}/(1-e^2)$ at small eccentricities~\cite{Peters:1964zz}.
The eccentricity for these channels thus decays less quickly than in vacuum. Nevertheless, even when additional massless fields are considered,
circular orbits remain stable.

By contrast, we show that sources of interest for GW detectors, evolving in thin accretion disks or other relatively large-density environment may inherit
a substantial eccentricity by the time they reach the mHz band. As we showed, high eccentricity is also a key feature of large mass ratio binaries, which is one possible explanation of
the GW190521 event~\cite{Nitz:2020mga}.
Together with previous results on the center-of-mass velocity of asymmetric binaries~\cite{Cardoso:2020lxx},
these results show that modeling binaries in accretion disks or nontrivial environments is challenging but crucial. In particular, these effects may have an important impact in attempts at constraining environmental properties~\cite{Barausse:2014tra,Cardoso:2019rou,Annulli:2020lyc,Toubiana:2020drf} or on testing fundamental properties of compact binaries~\cite{Cardoso:2019upw,Cardoso:2020nst}.

Our results complement previous findings~\cite{Roedig:2011rn,Zrake:2020zkw}. In particular, eccentricity excitation via asymmetric torques from circumbinary discs was found to keep supermassive black holes on eccentric orbits for a relevant fraction of their evolutionary phase~\cite{Roedig:2011rn}. Along the same line, it was recently shown that circumbinary disk torques may lead an equal-mass binary to evolve towards an equilibrium orbital eccentricity of~$e\simeq 0.45$~\cite{Zrake:2020zkw}. Interestingly, in that same analysis it was found that, when the circumbinary gas is in a thin disk, DF causes a {\it damping} in the eccentricity if the orbital eccentricity is~$e>0.45$. This effect is not captured by our model, as we do not consider the full modeling of the fluid perturbations and its gravitational effects.

%%%%%%%%%%%%%%%%%%%%%%%%%%%%%%%%%%%%%%%%%%%%%%%%%%%%%%%%%%%%%%%%%%%%%%%%%%%%%
\section*{Acknowledgements}
%%%%%%%%%%%%%%%%%%%%%%%%%%%%%%%%%%%%%%%%%%%%%%%%%%%%%%%%%%%%%%%%%%%%%%%%%%%%%
V.~C.\ acknowledges financial support provided under the European Union's H2020 ERC 
Consolidator Grant ``Matter and strong-field gravity: New frontiers in Einstein's 
theory'' grant agreement no. MaGRaTh--646597.
C.F.B.M acknowledges Conselho Nacional de
Desenvolvimento Científico e Tecnológico (CNPq), and
Coordenação de Aperfeiçoamento de Pessoal de Nível Superior (CAPES), from Brazil.
R.V.\ was  supported by the FCT PhD scholarship SFRH/BD/128834/2017.
This project has received funding from the European Union's Horizon 2020 research and innovation 
programme under the Marie Sklodowska-Curie grant agreement No 690904.
We thank FCT for financial support through Project~No.~UIDB/00099/2020.
We acknowledge financial support provided by FCT/Portugal through grant PTDC/MAT-APL/30043/2017.
The authors would like to acknowledge networking support by the GWverse COST Action 
CA16104, ``Black holes, gravitational waves and fundamental physics.''
%
%%%%%%%%%%%%%%%%%%%%%%%%%%%%%%%%%%%%%%%%%%%%%%%%%%%%%%%%%%%%%%%%%%%%%%%%%%%%%

\appendix
%%%%%%%%%%%%%%%%%%%%%%%%%%%%%%%%%%%%%%%%%%%%%%%%%%%%%%%%%%%%%%%%%
\section{Scalar and vector radiation\label{sec:details_emission}}
%%%%%%%%%%%%%%%%%%%%%%%%%%%%%%%%%%%%%%%%%%%%%%%%%%%%%%%%%%%%%%%%%

In addition to GW emission, many theories predict that binary could also emit through other channels, such as scalar and vector radiation. These additional emission can take place, for instance, if the BHs composing the binaries have scalar charges, as it is the case for self-interacting scalar fields, or even electromagnetic charges, as predicted by the Kerr-Newman class of BHs. In what follows, we explore the consequences of additional radiative sectors for the evolution of binaries.

%%%%%%%%%%%%%%%%%%%%%%%%%%
\subsection{Scalar charge}
%%%%%%%%%%%%%%%%%%%%%%%%%%
%%%%%%%%%%%%%%%%%%%%%%%%%%
\subsubsection{The theory}
%%%%%%%%%%%%%%%%%%%%%%%%%%
Consider the following theory describing a real massless scalar field~$\Phi$ sourced by two particles moving on a curved spacetime with metric~$g_{\mu \nu}$: 
\begin{align} \label{Action_Scalar}
	&\mathcal{S}=\int d^4x \sqrt{-g}\bigg[\frac{R}{8 \pi G}-g^{\mu \nu}\Phi_{,\mu} \Phi_ {,\nu}\nonumber\\
	&-\frac{2}{\sqrt{-g}}\sum_{j=1}^{2}(m_j+ 4 \pi q^0_j \Phi) \int d\lambda \sqrt{-g_{\mu \nu}\dot{z_j}^\mu \dot{z_j}^\nu}\delta^4(x-z_j)\bigg]\,,
\end{align}
with~$\Phi_{,\mu}\equiv \partial\Phi/\partial x^\mu$ and the determinant~$g\equiv {\rm det}(g_{\mu \nu})$. Here $z_j^\mu(\lambda)$ is the world line of the particle~$j=\{1,2\}$ parametrized by~$\lambda$, with~$\dot{z}_j^{\mu}\equiv d z_j^\mu/d\lambda$. Particle~$j$ has mass and scalar charge, respectively,~$m_j$ and $q^0_j$.
This theory has been extensively studied (see, e.g., Refs.~\cite{Burko:2002ge,Quinn:2000wa}).

Taking the variation of the action with respect to~$g_{\mu \nu}$ yields
\begin{align} \label{Einstein_EOM}
	G_{\mu \nu}=8 \pi G &\Bigg[T_{\mu \nu}^S+\frac{1}{\sqrt{-g}} \sum_{j=1}^{2}(m_j+ 4 \pi q^0_j \Phi)\nonumber\\ 
	&\times\int d\lambda \frac{(\dot{z}_j)_\mu (\dot{z}_j)_\nu}{\sqrt{-g_{\alpha \beta}\dot{z_j}^\alpha \dot{z_j}^\beta}}\delta^4(x-z_j(\lambda))\Bigg]\,,
\end{align}
with the scalar stress-energy tensor
\begin{align} \label{Scalar_SET}
	T_{\mu \nu}^S= \Phi_{,\mu} \Phi_ {,\nu}-\frac{g_{\mu \nu}}{2}g^{\alpha \beta} \Phi_{,\alpha} \Phi_{,\beta}\,.
\end{align}
The variation of~$\mathcal{S}$ with respect to~$\delta \Phi$ gives
\begin{align} 
	&\frac{1}{\sqrt{-g}}\partial_\mu\left(\sqrt{-g}g^{\mu \nu} \partial_\nu \Phi\right)=\nonumber\\
	&\frac{4\pi}{\sqrt{-g}}  \sum_{j=1}^{2}q^0_j\int d\lambda \sqrt{-g_{\mu \nu}\dot{z_j}^\mu \dot{z_j}^\nu}\delta^4(x-z_j(\lambda))\,,\label{Scalar_EOM}
\end{align}
and with respect to~$\delta z_j^\mu$ gives
\begin{align} 
	\left(m_j+4\pi q^0_j \Phi\right) u_j^\alpha \nabla_\alpha u_j^\mu=-4\pi q^0_j\left(g^{\mu \alpha}+u_j^\mu u_j^\alpha\right)\Phi_{,\alpha}\,,\label{Particle_EOM}
\end{align}
where~$\nabla$ is the Levi-Civita covariant derivative,~$u_j^\mu\equiv dz_j^\mu/d\tau_j$ is the 4-velocity of particle~$j$ and $\tau_j$ is its proper time.

%%%%%%%%%%%%%%%%%%%%%%%%%%%%%%%%%%%%%%%%%%%%%%%%%%
\subsubsection{Newtonian binary with no radiation}
%%%%%%%%%%%%%%%%%%%%%%%%%%%%%%%%%%%%%%%%%%%%%%%%%%
Consider a slowly-moving, Newtonian binary, such that energy and angular momentum fluxes can be neglected at leading order. In this limit~Eq.~\eqref{Einstein_EOM} becomes a simple Poisson equation~\cite{Poisson_will_2014}.
\begin{align}
\nabla^2 U=4\pi G \sum_{j=1}^2 m_j \delta^{3}(\boldsymbol{x}-\boldsymbol{r}_j(t))\,, \label{Poisson_eq}
\end{align}
where~$z_j^\mu\equiv(t,\boldsymbol{r}_j(t))$. The gravitational potential~$U(t,\boldsymbol{x})$ is weak, \textit{i.e.}~$|U|\ll1$, and enters in the Newtonian metric
\begin{align} 
ds^2=-(1+2 U)dt^2+dr^2+r^2\left(d \theta^2+ \sin^2\theta d\varphi^2\right)\,.\label{Newtonian_metric}
\end{align}
There is a (slowly time-varying) scalar field sourced by the point charges described by Eq.~\eqref{Scalar_EOM}, which in this limit becomes also a Poisson equation
\begin{align}
	\nabla^2 \Phi_0=4\pi \sum_{j=1}^2 q^0_j \delta^{3}(\boldsymbol{x}-\boldsymbol{r}_j(t))\,, \label{Poisson_eq1}
\end{align}
The equation of motion of the particles~\eqref{Particle_EOM} simplifies to a geodesic equation 
\begin{align}
	u_j^\alpha \nabla_\alpha u_j^\mu=-\frac{4 \pi q^0_j}{m_j+4\pi q^0_j \Phi_0}g^{\mu \alpha} \Phi_{,\alpha}\,.
\end{align}
We see that the particles are accelerated by the scalar. With the Newtonian metric~\eqref{Newtonian_metric} and assuming $q_1,q_2 \ll |\boldsymbol{r}_2-\boldsymbol{r}_1|$, this equation can be written in a familiar form~\footnote{One can see this directly by plugging the Newtonian metric~\eqref{Newtonian_metric} inside the particle's action in~\eqref{Action_Scalar}, obtaining 
\begin{align}
\mathcal{S}_{\rm part}&=\sum_ j m_j\int dt \sqrt{(1+2 U)-|d\boldsymbol{r}_j/dt|^2} \nonumber \\
&\simeq \sum_ j m_j\int dt\left( 1+U-\tfrac{1}{2}|d\boldsymbol{r}_j/dt|^2\right)\,.
\end{align}
This is just the action describing a non-relativistic system of particles in a gravitational potential~$U$.
}
\begin{align}\label{Newton_2law}
	\frac{d^2}{d t^2} \boldsymbol{r}_j=-\boldsymbol{\nabla} U(t,\boldsymbol{r}_j)-4\pi \frac{q^0_j}{m_ j} \boldsymbol{\nabla} \Phi_0(t,\boldsymbol{r}_j)\,,
\end{align}
where $\boldsymbol{\nabla}$ is the usual $3$-dimensional gradient operator. 
Using equation~\eqref{Poisson_eq} we obtain~\footnote{Actually, in this step we cannot really consider point sources, otherwise we would find problems with a diverging ``self-force''. Fortunately, this is not a real problem, and we can proceed by assuming that the particles have a small, but finite, size.}
\begin{align}
	&U(t,\boldsymbol{r}_1)=\frac{G m_2}{|\boldsymbol{r}_2(t)-\boldsymbol{r}_1|}\,, \quad &U(t,\boldsymbol{r}_2)=\frac{G m_1}{|\boldsymbol{r}_2-\boldsymbol{r}_1(t)|}\,, \\
	&\Phi_0(t,\boldsymbol{r}_1)=\frac{q_2^0}{|\boldsymbol{r}_2(t)-\boldsymbol{r}_1|}\,, \quad &\Phi_0(t,\boldsymbol{r}_2)=\frac{q_1^0}{|\boldsymbol{r}_2-\boldsymbol{r}_1(t)|}\,,\label{Coloumb_pot}
\end{align}
%
%%%%%%%%%%%%%%%%%%%%%%%%%%%%%%%%%%%%%%%%%%%%%%%%%%
\subsubsection{Elliptic motion and orbit-averaging}
%%%%%%%%%%%%%%%%%%%%%%%%%%%%%%%%%%%%%%%%%%%%%%%%%%
As one expects, Eq.~\eqref{Newton_2law} with~\eqref{Coloumb_pot} describes the Keplerian orbital motion with energy and angular momentum given in Eq.~\eqref{eq:energy_angularmomentum_main}.
%
%\begin{align}
	%E&=-\frac{\tilde{G} m_1m_2}{2a}\,,\\
	%%
	%L^2&=\frac{\tilde{G}m_1^2m_2^2a(1-e^2)}{M}\,,
%\end{align}
%%
%where
%%
%\begin{align}
	%\tilde{G}\equiv 1-4 \pi \frac{q_1 q_2}{m_1 m_ 2}\,.
%\end{align}
%
These differ from~\eqref{Kepler_Energy} and~\eqref{Kepler_AngMom} due to the scalar interaction. Using spherical coordinates with origin at the center of mass the trajectories can be written as~$\boldsymbol{r}_1=\left(r_1(\varphi_p),\varphi_p,\pi/2\right)$ and~$\boldsymbol{r}_2=\left(r_2(\varphi_p),\varphi_p+\pi,\pi/2\right)$ with
\begin{align}
r_1=\frac{m_2}{M}r_p \,, \qquad r_2=\frac{m_1}{M} r_p\,, 
\end{align}
\begin{align}
r_p(\varphi_p)=\frac{a(1-e^2)}{1+e \cos \varphi_p} \,.
\end{align}
Their angular velocity is
\begin{align}
	\dot{\varphi_p}=\sqrt{\frac{\tilde{G} M}{a^3}}(1-e^2)^{-3/2}(1+e\cos\varphi)^2\,.
\end{align}
%

%%
%\be
%v=\sqrt{\frac{\tilde{G} M(1+e^2+2e\cos\varphi)}{a(1-e^2)}}\,.
%\ee
%%

Finally, we define the average of a quantity $X$ over one period $T$ as
\begin{equation}
\left<X\right>=\frac{\omega_0}{2\pi}\int_0^{2\pi}\frac{d\varphi}{\dot{\varphi}}X(\varphi)\,.
\end{equation}
where $\omega_0$ is the (Keplerian) orbital frequency.
%%%%%%%%%%%%%%%%%%%%%%%%%%%%%%%%%%%%%%%%%%%%%%%%%%%%%%%%
\subsubsection{Radiation emitted by a Newtonian binary}
%%%%%%%%%%%%%%%%%%%%%%%%%%%%%%%%%%%%%%%%%%%%%%%%%%%%%%%%
A Newtonian binary sources a scalar field described by Eq.~\eqref{Scalar_EOM}, which can be put in the form
\begin{align} \label{Scalar_EOM1}
	\Box \Phi=4 \pi \rho(t,\boldsymbol{x}) \equiv\frac{4 \pi}{\sqrt{-g}} \sum_{j=1}^2 q^0_j\delta^3\left(\boldsymbol{x}-\boldsymbol{r}_j(t)\right)\,.
\end{align}
Thus, the binary will lose energy and angular momentum through this channel and the motion will not be truly Keplerian; the radiation reaction force entering~\eqref{Particle_EOM} (which we are neglecting in the computation of the radiation, because we are using an adiabatic approximation) will be responsible for a deviation to the Keplerian orbit. Let us compute the radiation emitted by this binary of scalar charges in the (leading) dipole approximation.

In the Newtonian approximation the scalar radiation propagates in flat space. So, the solution of (sourced) scalar wave equation is
\begin{align}
	\Phi(t, \boldsymbol{x})=\int d^3 \boldsymbol{x}'\sqrt{-g'}\, \frac{\rho(t-|\boldsymbol{x}-\boldsymbol{x}'|,\boldsymbol{x}')}{|\boldsymbol{x}-\boldsymbol{x}'|}\,.
\end{align}
In the dipole approximation it is easy to see that
\begin{align}
	\Phi(t,r\to \infty, \theta, \varphi)\simeq \frac{1}{r}\boldsymbol{e}_ r\cdot \dot{\boldsymbol{p}}(t-r)\,, 
\end{align}
with the dipole moment
\be
\boldsymbol{p}(t)\equiv \int d^3 \boldsymbol{x}' \sqrt{-g'} \rho(t, \boldsymbol{x}')\, \boldsymbol{x}'=\left(\frac{q^0_1 m_2-q^0_2 m_1}{M}\right) \boldsymbol{r}_p(t)\,.\nonumber
\ee
This approximation is valid for scalar waves with frequency~$\omega\sim \omega_0 \ll 1/a$, where~$\omega_0$ is the orbital frequency (which is compatible with the Newtonian approximation).
%
%Now we have all we need to compute the energy and angular momentum radiated with the scalar field. 
The radiated energy flux is
\begin{align}
\dot{E}^{\rm rad}=-\lim_{r \to \infty} r^2 \int d \Omega\, T^S_ {rt}\,,
\end{align}
and the angular momentum through
\begin{align}
\dot{L}^{\rm rad}=\lim_{r \to \infty} r^2 \int d \Omega\, T^S_ {r\varphi}\,.
\end{align}
Plugging the dipole approximation in the scalar's stress-energy tensor~\eqref{Scalar_SET} we can write the last two expressions in the form
\begin{align}
	\dot{E}^{\rm rad}&=\left(\frac{q^0_1 m_2-q^0_2 m_1}{M}\right)^2 \int d\Omega \, \left[\boldsymbol{e}_ r\cdot \ddot{\boldsymbol{r}}_p \right]^2 \nonumber \\
	&=\frac{4\pi}{3}\frac{\tilde{G}^2}{r_p^4}(q^0_1 m_2-q^0_2 m_1)^2\,, \label{EnergyS}
\end{align}
where we used~$\ddot{\boldsymbol{r}}_p=-\tilde{G} M \boldsymbol{r}_p/r_p^3$ and integrated over the sphere, and
\begin{align}
&\dot{L}^{\rm rad}=-\left(\frac{q^0_1 m_2-q^0_2 m_1}{M}\right)^2 \int d\Omega \, \left(\boldsymbol{e}_ r\cdot \ddot{\boldsymbol{r}}_p \right)\partial_\varphi \left(\boldsymbol{e}_ r\cdot \dot{\boldsymbol{r}}_p \right) \nonumber \\
&=\frac{4 \pi}{3} \tilde{G}^{\frac{3}{2}}\frac{\sqrt{a (1-e^2)}}{\sqrt{M}r_p^3}(q^0_1 m_2-q^0_2 m_1)^2\,.\label{AngMomS}
\end{align}
Averaging over an orbit we find
\begin{align}
&\langle\dot{E}^{\rm rad}\rangle=\frac{2\pi}{3}\frac{\tilde{G}^2}{a^4}(q^0_1 m_2-q^0_2 m_1)^2 \left(\frac{2+e^2}{(1-e^2)^{\frac{5}{2}}}\right) \label{EnergySav} \,, \\
&\langle\dot{L}^{\rm rad}\rangle=\frac{4 \pi}{3} \frac{\tilde{G}^\frac{3}{2}}{\sqrt{M}a^{\frac{5}{2}}(1-e^2)}(q^0_1 m_2-q^0_2 m_1)^2  \label{AngMomSav}\,,
\end{align}
resulting in the ratio
\begin{align}\label{LEratioS}
&\frac{\langle\dot{L}^{\rm rad} \rangle}{\langle\dot{E}^{\rm rad} \rangle} = \frac{\sqrt{1-e^2}}{\omega_0} \left(\frac{1-e^2}{1+\frac{e^2}{2}}\right) \,.
\end{align}
In the adiabatic approximation the major semi-axis and the eccentricity follow
\begin{align}
	\langle\dot{a}\rangle&=-\frac{2a^2\langle\dot{E}^{\rm rad}\rangle}{\tilde{G} m_1m_2}<0\,,\label{shrink}\\
	\langle\dot{e}\rangle&=\sqrt{\frac{M}{ \tilde{G}a}}\frac{\sqrt{1-e^2}}{e}\frac{\langle\dot{E}^{\rm rad}\rangle}{m_1m_2}\left(\frac{\langle\dot{L}^{\rm rad}\rangle}{\langle\dot{E}^{\rm rad}\rangle}-\frac{\sqrt{1-e^2}}{\omega_0}\right) \nonumber \\
	&=-\sqrt{\frac{M}{ \tilde{G}a}}\left(\frac{1-e^2}{e\, \omega_0}\right)\frac{\langle\dot{E}^{\rm rad}\rangle}{m_1m_2}\left(\frac{3 e^2}{2+e^2}\right)\leq 0\,.\label{circular}
\end{align}
Thus, the emission of scalar radiation by a binary causes the major semi-axis and the eccentricity to decrease in time: the orbit shrinks and circularizes. 
In the circular orbit limit our results are in agreement with those of Refs.~\cite{Cardoso:2011xi,Yunes:2011aa,Cardoso:2019nis}.
%%%%%%%%%%%%%%%%%%%%%%%%%%%%%%%%%%%%%%%%%%%%%%%%%%%%%%%%%%%%%%%%%%%%%%%%%%%%%%%
\subsection{Electric charge}
%%%%%%%%%%%%%%%%%%%%%%%%%%%%%%%%%%%%%%%%%%%%%%%%%%%%%%%%%%%%%%%%%%%%%%%%%%%%%%%
%%%%%%%%%%%%%%%%%%%%%
\subsubsection{Theory}
%%%%%%%%%%%%%%%%%%%%%
Here we consider the theory of an electromagnetic field~$A_\mu$ sourced by two electric charges moving on a curved spacetime with metric~$g_ {\mu \nu}$,
\begin{align} \label{Action_Vector}
\mathcal{S}&=\int d^4x \sqrt{-g}\bigg[\frac{R}{8 \pi}-\frac{1}{2}F^{\mu\nu} F_ {\mu \nu}\nonumber \\
&-\frac{2}{\sqrt{-g}}\sum_{j=1}^{2}m_j \int d\lambda \sqrt{-g_{\mu \nu}\dot{z_j}^\mu \dot{z_j}^\nu}\delta^4(x-z_j)\nonumber\\
&-\frac{8\pi}{\sqrt{-g}}\sum_{j=1}^2 q^1_j A_\alpha \int d\lambda\, \dot{z}_j^\alpha\delta^4(x-z_j)\bigg]\,,
\end{align}
where~$F_{\mu \nu}\equiv \partial_\mu A_\nu-\partial_\nu A_\mu$ and~$q^1_j$ is the electric charge of particle~$j$.

Taking the variation of the action with respect to~$A_\mu$ yields the (sourced) Maxwell equations
\begin{align}
&\partial_\mu F^{\mu \nu}=4 \pi J^\nu\,, \\
&J^\nu\equiv \frac{1}{\sqrt{-g}}\sum_{j=1}^2q^1_j u_j^\nu\delta^3\left(\boldsymbol{x}-\boldsymbol{r}_ j\right)\,,
\end{align} 
where~$u_j$ is the 4-velocity of particle~$j$. In the Newtonian approximation and neglecting radiation (valid for slowly moving charges) we can repeat the exact same steps that we applied to the scalar charges to find that the electric charges also describe a Keplerian orbit; the only difference being that in the definition of~$\tilde{G}$ we have now electric charges instead of scalar charges. 

The stress-energy tensor of the electromagnetic field is
\begin{align}
	T_{\mu \nu}^{EM}=-\frac{1}{4}F^{\alpha \beta}F_{\alpha \beta} g_{\mu\nu}+F_ {\mu \alpha}F_\nu^{\;\,\alpha}\,.
\end{align}
%
%%%%%%%%%%%%%%%%%%%%%%%%%%%%%%%%%%%%%%%%%%%%%%%%%%%%%%%
\subsubsection{Radiation emitted by a Newtonian binary}
%%%%%%%%%%%%%%%%%%%%%%%%%%%%%%%%%%%%%%%%%%%%%%%%%%%%%%%
Again, the binary will radiate energy and angular momentum -- in this case through electromagnetic waves -- and the motion will not be truly Keplerian; in the regime we are considering, the orbits will change adiabatically.

Using the Lorenz gauge~$\partial_\mu A^\mu=0$ the sourced Maxwell equations become
\begin{align}
\Box A^\alpha=4 \pi J^\alpha\,,
\end{align}
which we can decompose into
\begin{align}
&\Box \Phi=4 \pi \rho(t,\boldsymbol{x})\equiv\frac{4\pi}{\sqrt{-g}}\sum_{j=1}^2q^1_j \delta^3\left(\boldsymbol{x}-\boldsymbol{r}_ j\right)\,, \\
&\Box \boldsymbol{A}= 4\pi \boldsymbol{j}(t, \boldsymbol{x})\equiv\frac{4\pi}{\sqrt{-g}}\sum_{j=1}^2q^1_j \boldsymbol{v}_j \delta^3\left(\boldsymbol{x}-\boldsymbol{r}_ j\right)\,,
\end{align}
where we used that the sources are non-relativistic. In the Newtonian approximation we consider that the electromagnetic waves propagate in flat space. So, the solution to the (sourced) Maxwell equations is
\begin{align}
&\Phi(t, \boldsymbol{x})=\int d^3 \boldsymbol{x}'\sqrt{-g'}\, \frac{\rho(t-|\boldsymbol{x}-\boldsymbol{x}'|,\boldsymbol{x}')}{|\boldsymbol{x}-\boldsymbol{x}'|}\,, \\
&\boldsymbol{A}(t, \boldsymbol{x})=\int d^3 \boldsymbol{x}'\sqrt{-g'}\, \frac{\boldsymbol{j}(t-|\boldsymbol{x}-\boldsymbol{x}'|,\boldsymbol{x}')}{|\boldsymbol{x}-\boldsymbol{x}'|}\,.
\end{align}
In the dipole approximation one can show that
\begin{align}
&\Phi(t,r\to \infty, \theta, \varphi)\simeq \frac{1}{r}\boldsymbol{e}_ r\cdot \dot{\boldsymbol{p}}(t-r)\,, \\
&\boldsymbol{A}(t,r\to \infty, \theta, \varphi)\simeq \frac{1}{r} \dot{\boldsymbol{p}}(t-r)\,,
\end{align}
with the dipole moment
\begin{align}
\boldsymbol{p}(t)\equiv \int d^3 \boldsymbol{x}' \sqrt{-g'} \rho(t, \boldsymbol{x}')\, \boldsymbol{x}' =\left(\frac{q^1_1 m_2-q^1_2 m_1}{M}\right) \boldsymbol{r}_p(t)\,.\nonumber
\end{align}
Now, the magnetic field is
\begin{align}
	\boldsymbol{B}(t, r\to\infty,\theta,\varphi)\equiv \boldsymbol{\nabla} \times \boldsymbol{A}\simeq -\frac{1}{r} \boldsymbol{e}_r \times \ddot{\boldsymbol{p}}(t-r)
\end{align}
and using Ampère-Maxwell's law we have
\begin{align}
	\dot{\boldsymbol{E}}(t, r\to \infty,\theta,\varphi)=\boldsymbol{\nabla}\times \boldsymbol{B}=\dot{\boldsymbol{B}}\times\boldsymbol{e}_ r\,,
\end{align}
which, integrating in time, gives the electric field
\begin{align}
	\boldsymbol{E}(t, r\to \infty,\theta,\varphi)=\boldsymbol{B}\times\boldsymbol{e}_ r\,.
\end{align}
These result in the Poynting vector
\begin{align}
	\boldsymbol{S}(t,r\to \infty,\theta,\varphi)\equiv \boldsymbol{E}\times \boldsymbol{B}=|\boldsymbol{B}|^2\boldsymbol{e}_r\,, 
\end{align}
where we used Lagrange's rule for the triple cross product and that~$(\boldsymbol{B}\cdot \boldsymbol{e}_r)=0$.
Now using the scalar quadruple product identity we have
\begin{align}
	&|\boldsymbol{B}|^2=\frac{1}{r^2}\left(|\ddot{\boldsymbol{p}}|^2-\left(\ddot{\boldsymbol{p}}\cdot \boldsymbol{e}_ r\right)^2\right) \,.
\end{align}
So the radiated energy flux is
\begin{align}
	\dot{E}^{\rm rad}&=-\lim_{r \to \infty} r^2 \int d \Omega\, T^{EM}_ {rt}=\lim_{r \to \infty} r^2 \int d\Omega\, \boldsymbol{S}\cdot \boldsymbol{e}_r \nonumber \\
	&=\left(\frac{q^1_1 m_2-q^1_2 m_1}{M}\right)^2 \int d\Omega \, \left[|\ddot{\boldsymbol{r}}_p|^2-\left(\ddot{\boldsymbol{r}}_p\cdot \boldsymbol{e}_ r \right)^2\right] \nonumber \\
	&=\frac{8\pi}{3}\frac{\tilde{G}^2}{r_p^4}(q^1_1 m_2-q^1_2 m_1)^2\,,
\end{align}
where we used~$\ddot{\boldsymbol{r}}_p=-\tilde{G} M \boldsymbol{e}_ r/r^2$ and integrated over the sphere.
The radiated angular momentum flux
\begin{align}
		\dot{L}^{\rm rad}&=\lim_{r \to \infty} r^2 \int d \Omega\, T^{EM}_ {r\varphi}\nonumber \\
		&=2 \left(\frac{q^1_1 m_2-q^1_2m_1}{M}\right)^2 \nonumber\\ 
		&\times\int d\Omega \, \left[\left(\boldsymbol{e}_ r\cdot \ddot{\boldsymbol{r}}_p \right)\partial_\varphi \left(\boldsymbol{e}_ r\cdot \dot{\boldsymbol{r}}_p \right)-\left(\boldsymbol{e}_ r\cdot \ddot{\boldsymbol{r}}_p \right)\left(\boldsymbol{e}_ \varphi\cdot \ddot{\boldsymbol{r}}_p \right)\right]\nonumber \\
		&=\frac{8 \pi}{3} \tilde{G}^\frac{3}{2}\frac{\sqrt{a (1-e^2)}}{\sqrt{M}r_p^3}(q^1_1 m_2-q^1_2 m_1)^2\,.
\end{align}
Thus, averaging over one orbital period, we conclude that the electric charges radiate twice the energy and twice the angular momentum per unit of time in comparison with the scalar charges (compare with Eqs.~\eqref{EnergyS} and~\eqref{AngMomS}). So, the ratio between the angular momentum and energy carried by the radiated electromagnetic field~$\langle\dot{L}^{\rm rad} \rangle/\langle\dot{E}^{\rm rad} \rangle$ is the same as for the scalar field and is given by~\eqref{LEratioS}. So, the emission of electromagnetic waves by a binary causes both the major semi-axis and eccentricity to decrease in time: the orbit shrinks and circularizes (see~\eqref{shrink} and~\eqref{circular}). 
Our results for the electromagnetic radiation emitted by a binary are in agreement with the ones of Refs.~\cite{Christiansen:2020pnv,Liu:2020cds}.

%%%%%%%%%%%%%%%%%%%%%%%%%%%%%%%%%%%%%%%%%%%%%%%%%%%%%%%%%%%%%%%%%%%%%
\section{When the adiabatic assumption fails}\label{sec:adiabatic}
%%%%%%%%%%%%%%%%%%%%%%%%%%%%%%%%%%%%%%%%%%%%%%%%%%%%%%%%%%%%%%%%%%%%%
We have made extensive use of the adiabatic approximation in the main text to analyze the evolution of the eccentricity of the system subjected to the GW and environmental forces. However, depending on the environmental density and the initial separation of the binary, this approximation may not be valid. In this subsection, we address how much the adiabatic approximation may underestimate the eccentricity increase in the system.
In order to investigate the validity of the adiabatic approximation for equal mass binaries, we integrate Eq.~\eqref{eq:r_eq} (neglecting accretion), considering specific initial conditions. With the numerical solution, we construct the eccentricity as function of the orbital distance, by tracking the expression~\eqref{eq:ecc}. Since this system only takes into account the environmental effects, we compare this solution to the one obtained from the adiabatic approach by integrating Eq.~\eqref{eq:dfecc} under similar conditions (with higher order of eccentricity included). With the results, we compute the relative deviation of the eccentricity, \textit{i.e},
\begin{equation}
\frac{\delta e}{e_a}=\frac{|e_n-e_a|}{e_a}\,,
\end{equation}
where $e_n$ is the result from Eq.~\eqref{eq:r_eq} and $e_a$ the one from the adiabatic approximation (considering terms up to ${\cal O}(e^{12})$). The deviation depends on the medium density and the initial conditions, but we expect it to approach zero as the medium density decreases.

\begin{figure}[ht]
	\includegraphics[width=\linewidth]{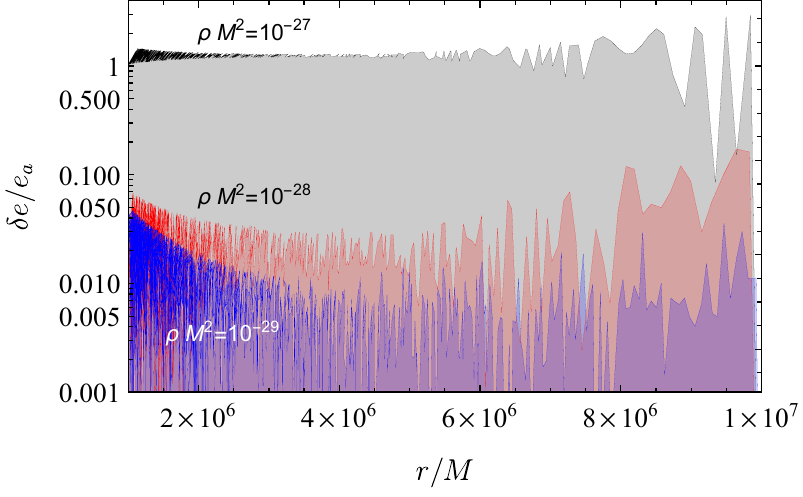}
	\caption{Comparison between the numerical integration of Eq.~\eqref{eq:r_eq} and the result from the adiabatic approach. We plot the deviation normalized by the adiabatic result.}
	\label{fig:deviation}
\end{figure}

In Fig.~\ref{fig:deviation} we plot the eccentricity deviation, considering initial separation of $a=10^7M$ and initial eccentricity $e_0=0.001$. For the dynamical friction, we consider $\lambda=20$. We can see that for densities of $\rho M^2=10^{-27}$ the adiabatic approximation fails to quantitatively describe the eccentricity evolution of the system, underestimating the eccentricity increasing from the DF. For densities as small as $\rho M^2=10^{-29}$ the adiabatic approach works mostly in the initial stages of the binary evolution. At late times, meaning short distances, we can see that the eccentricity deviation increases, indicating a possible breaking of the adiabatic approximation.

The discrepancy between the adiabatic and the numerical computation of the eccentricity increases at late times (smaller orbital distances), showing that we cannot underestimate the contribution from the environmental forces. Going beyond the adiabatic approximation shows that the eccentricity increases even further; this effect is enhanced for asymmetric binaries and accretion, as we discussed in the main text.

\bibliography{References}

\begin{thebibliography}{54}%
\makeatletter
\providecommand \@ifxundefined [1]{%
 \@ifx{#1\undefined}
}%
\providecommand \@ifnum [1]{%
 \ifnum #1\expandafter \@firstoftwo
 \else \expandafter \@secondoftwo
 \fi
}%
\providecommand \@ifx [1]{%
 \ifx #1\expandafter \@firstoftwo
 \else \expandafter \@secondoftwo
 \fi
}%
\providecommand \natexlab [1]{#1}%
\providecommand \enquote  [1]{``#1''}%
\providecommand \bibnamefont  [1]{#1}%
\providecommand \bibfnamefont [1]{#1}%
\providecommand \citenamefont [1]{#1}%
\providecommand \href@noop [0]{\@secondoftwo}%
\providecommand \href [0]{\begingroup \@sanitize@url \@href}%
\providecommand \@href[1]{\@@startlink{#1}\@@href}%
\providecommand \@@href[1]{\endgroup#1\@@endlink}%
\providecommand \@sanitize@url [0]{\catcode `\\12\catcode `\$12\catcode
  `\&12\catcode `\#12\catcode `\^12\catcode `\_12\catcode `\%12\relax}%
\providecommand \@@startlink[1]{}%
\providecommand \@@endlink[0]{}%
\providecommand \url  [0]{\begingroup\@sanitize@url \@url }%
\providecommand \@url [1]{\endgroup\@href {#1}{\urlprefix }}%
\providecommand \urlprefix  [0]{URL }%
\providecommand \Eprint [0]{\href }%
\providecommand \doibase [0]{http://dx.doi.org/}%
\providecommand \selectlanguage [0]{\@gobble}%
\providecommand \bibinfo  [0]{\@secondoftwo}%
\providecommand \bibfield  [0]{\@secondoftwo}%
\providecommand \translation [1]{[#1]}%
\providecommand \BibitemOpen [0]{}%
\providecommand \bibitemStop [0]{}%
\providecommand \bibitemNoStop [0]{.\EOS\space}%
\providecommand \EOS [0]{\spacefactor3000\relax}%
\providecommand \BibitemShut  [1]{\csname bibitem#1\endcsname}%
\let\auto@bib@innerbib\@empty
%</preamble>
\bibitem [{\citenamefont {Abbott}\ \emph {et~al.}(2016)\citenamefont {Abbott}
  \emph {et~al.}}]{Abbott:2016blz}%
  \BibitemOpen
  \bibfield  {author} {\bibinfo {author} {\bibfnamefont {B.}~\bibnamefont
  {Abbott}} \emph {et~al.} (\bibinfo {collaboration} {LIGO Scientific,
  Virgo}),\ }\href {\doibase 10.1103/PhysRevLett.116.061102} {\bibfield
  {journal} {\bibinfo  {journal} {Phys. Rev. Lett.}\ }\textbf {\bibinfo
  {volume} {116}},\ \bibinfo {pages} {061102} (\bibinfo {year} {2016})},\
  \Eprint {http://arxiv.org/abs/1602.03837} {arXiv:1602.03837 [gr-qc]}
  \BibitemShut {NoStop}%
\bibitem [{\citenamefont {Barack}\ \emph {et~al.}(2019)\citenamefont {Barack}
  \emph {et~al.}}]{Barack:2018yly}%
  \BibitemOpen
  \bibfield  {author} {\bibinfo {author} {\bibfnamefont {L.}~\bibnamefont
  {Barack}} \emph {et~al.},\ }\href {\doibase 10.1088/1361-6382/ab0587}
  {\bibfield  {journal} {\bibinfo  {journal} {Class. Quant. Grav.}\ }\textbf
  {\bibinfo {volume} {36}},\ \bibinfo {pages} {143001} (\bibinfo {year}
  {2019})},\ \Eprint {http://arxiv.org/abs/1806.05195} {arXiv:1806.05195
  [gr-qc]} \BibitemShut {NoStop}%
\bibitem [{\citenamefont {Peters}(1964)}]{Peters:1964zz}%
  \BibitemOpen
  \bibfield  {author} {\bibinfo {author} {\bibfnamefont {P.}~\bibnamefont
  {Peters}},\ }\href {\doibase 10.1103/PhysRev.136.B1224} {\bibfield  {journal}
  {\bibinfo  {journal} {Phys. Rev.}\ }\textbf {\bibinfo {volume} {136}},\
  \bibinfo {pages} {B1224} (\bibinfo {year} {1964})}\BibitemShut {NoStop}%
\bibitem [{\citenamefont {Krolak}\ and\ \citenamefont
  {Schutz}(1987)}]{Krolak:1987ofj}%
  \BibitemOpen
  \bibfield  {author} {\bibinfo {author} {\bibfnamefont {A.}~\bibnamefont
  {Krolak}}\ and\ \bibinfo {author} {\bibfnamefont {B.~F.}\ \bibnamefont
  {Schutz}},\ }\href {\doibase 10.1007/BF00759095} {\bibfield  {journal}
  {\bibinfo  {journal} {Gen. Rel. Grav.}\ }\textbf {\bibinfo {volume} {19}},\
  \bibinfo {pages} {1163} (\bibinfo {year} {1987})}\BibitemShut {NoStop}%
\bibitem [{\citenamefont {Shapiro~Key}\ and\ \citenamefont
  {Cornish}(2011)}]{Key:2010tc}%
  \BibitemOpen
  \bibfield  {author} {\bibinfo {author} {\bibfnamefont {J.}~\bibnamefont
  {Shapiro~Key}}\ and\ \bibinfo {author} {\bibfnamefont {N.~J.}\ \bibnamefont
  {Cornish}},\ }\href {\doibase 10.1103/PhysRevD.83.083001} {\bibfield
  {journal} {\bibinfo  {journal} {Phys. Rev. D}\ }\textbf {\bibinfo {volume}
  {83}},\ \bibinfo {pages} {083001} (\bibinfo {year} {2011})},\ \Eprint
  {http://arxiv.org/abs/1006.3759} {arXiv:1006.3759 [gr-qc]} \BibitemShut
  {NoStop}%
\bibitem [{\citenamefont {Laine}\ \emph {et~al.}(2020)\citenamefont {Laine}
  \emph {et~al.}}]{Laine:2020dnr}%
  \BibitemOpen
  \bibfield  {author} {\bibinfo {author} {\bibfnamefont {S.}~\bibnamefont
  {Laine}} \emph {et~al.},\ }\href {\doibase 10.3847/2041-8213/ab79a4}
  {\bibfield  {journal} {\bibinfo  {journal} {Astrophys. J. Lett.}\ }\textbf
  {\bibinfo {volume} {894}},\ \bibinfo {pages} {L1} (\bibinfo {year} {2020})},\
  \Eprint {http://arxiv.org/abs/2004.13392} {arXiv:2004.13392 [astro-ph.HE]}
  \BibitemShut {NoStop}%
\bibitem [{\citenamefont {Abbott}\ \emph
  {et~al.}(2020{\natexlab{a}})\citenamefont {Abbott} \emph
  {et~al.}}]{Abbott:2020tfl}%
  \BibitemOpen
  \bibfield  {author} {\bibinfo {author} {\bibfnamefont {R.}~\bibnamefont
  {Abbott}} \emph {et~al.} (\bibinfo {collaboration} {LIGO Scientific,
  Virgo}),\ }\href {\doibase 10.1103/PhysRevLett.125.101102} {\bibfield
  {journal} {\bibinfo  {journal} {Phys. Rev. Lett.}\ }\textbf {\bibinfo
  {volume} {125}},\ \bibinfo {pages} {101102} (\bibinfo {year}
  {2020}{\natexlab{a}})},\ \Eprint {http://arxiv.org/abs/2009.01075}
  {arXiv:2009.01075 [gr-qc]} \BibitemShut {NoStop}%
\bibitem [{\citenamefont {Abbott}\ \emph
  {et~al.}(2020{\natexlab{b}})\citenamefont {Abbott} \emph
  {et~al.}}]{Abbott:2020mjq}%
  \BibitemOpen
  \bibfield  {author} {\bibinfo {author} {\bibfnamefont {R.}~\bibnamefont
  {Abbott}} \emph {et~al.} (\bibinfo {collaboration} {LIGO Scientific,
  Virgo}),\ }\href {\doibase 10.3847/2041-8213/aba493} {\bibfield  {journal}
  {\bibinfo  {journal} {Astrophys. J.}\ }\textbf {\bibinfo {volume} {900}},\
  \bibinfo {pages} {L13} (\bibinfo {year} {2020}{\natexlab{b}})},\ \Eprint
  {http://arxiv.org/abs/2009.01190} {arXiv:2009.01190 [astro-ph.HE]}
  \BibitemShut {NoStop}%
\bibitem [{\citenamefont {Gayathri}\ \emph {et~al.}(2020)\citenamefont
  {Gayathri}, \citenamefont {Healy}, \citenamefont {Lange}, \citenamefont
  {O'Brien}, \citenamefont {Szczepanczyk}, \citenamefont {Bartos},
  \citenamefont {Campanelli}, \citenamefont {Klimenko}, \citenamefont
  {Lousto},\ and\ \citenamefont {O'Shaughnessy}}]{Gayathri:2020coq}%
  \BibitemOpen
  \bibfield  {author} {\bibinfo {author} {\bibfnamefont {V.}~\bibnamefont
  {Gayathri}}, \bibinfo {author} {\bibfnamefont {J.}~\bibnamefont {Healy}},
  \bibinfo {author} {\bibfnamefont {J.}~\bibnamefont {Lange}}, \bibinfo
  {author} {\bibfnamefont {B.}~\bibnamefont {O'Brien}}, \bibinfo {author}
  {\bibfnamefont {M.}~\bibnamefont {Szczepanczyk}}, \bibinfo {author}
  {\bibfnamefont {I.}~\bibnamefont {Bartos}}, \bibinfo {author} {\bibfnamefont
  {M.}~\bibnamefont {Campanelli}}, \bibinfo {author} {\bibfnamefont
  {S.}~\bibnamefont {Klimenko}}, \bibinfo {author} {\bibfnamefont
  {C.}~\bibnamefont {Lousto}}, \ and\ \bibinfo {author} {\bibfnamefont
  {R.}~\bibnamefont {O'Shaughnessy}},\ }\href@noop {} {\  (\bibinfo {year}
  {2020})},\ \Eprint {http://arxiv.org/abs/2009.05461} {arXiv:2009.05461
  [astro-ph.HE]} \BibitemShut {NoStop}%
\bibitem [{\citenamefont {Calder\'on~Bustillo}\ \emph
  {et~al.}(2020)\citenamefont {Calder\'on~Bustillo}, \citenamefont
  {Sanchis-Gual}, \citenamefont {Torres-Forn\'e},\ and\ \citenamefont
  {Font}}]{CalderonBustillo:2020odh}%
  \BibitemOpen
  \bibfield  {author} {\bibinfo {author} {\bibfnamefont {J.}~\bibnamefont
  {Calder\'on~Bustillo}}, \bibinfo {author} {\bibfnamefont {N.}~\bibnamefont
  {Sanchis-Gual}}, \bibinfo {author} {\bibfnamefont {A.}~\bibnamefont
  {Torres-Forn\'e}}, \ and\ \bibinfo {author} {\bibfnamefont {J.~A.}\
  \bibnamefont {Font}},\ }\href@noop {} {\  (\bibinfo {year} {2020})},\ \Eprint
  {http://arxiv.org/abs/2009.01066} {arXiv:2009.01066 [gr-qc]} \BibitemShut
  {NoStop}%
\bibitem [{\citenamefont {Graham}\ \emph {et~al.}(2020)\citenamefont {Graham}
  \emph {et~al.}}]{Graham:2020gwr}%
  \BibitemOpen
  \bibfield  {author} {\bibinfo {author} {\bibfnamefont {M.}~\bibnamefont
  {Graham}} \emph {et~al.},\ }\href {\doibase 10.1103/PhysRevLett.124.251102}
  {\bibfield  {journal} {\bibinfo  {journal} {Phys. Rev. Lett.}\ }\textbf
  {\bibinfo {volume} {124}},\ \bibinfo {pages} {251102} (\bibinfo {year}
  {2020})},\ \Eprint {http://arxiv.org/abs/2006.14122} {arXiv:2006.14122
  [astro-ph.HE]} \BibitemShut {NoStop}%
\bibitem [{\citenamefont {Cardoso}\ and\ \citenamefont
  {Macedo}(2020)}]{Cardoso:2020lxx}%
  \BibitemOpen
  \bibfield  {author} {\bibinfo {author} {\bibfnamefont {V.}~\bibnamefont
  {Cardoso}}\ and\ \bibinfo {author} {\bibfnamefont {C.~F.}\ \bibnamefont
  {Macedo}},\ }\href {\doibase 10.1093/mnras/staa2396} {\  (\bibinfo {year}
  {2020}),\ 10.1093/mnras/staa2396},\ \Eprint {http://arxiv.org/abs/2008.01091}
  {arXiv:2008.01091 [astro-ph.HE]} \BibitemShut {NoStop}%
\bibitem [{\citenamefont {Gergely}\ \emph {et~al.}(1998)\citenamefont
  {Gergely}, \citenamefont {Perjes},\ and\ \citenamefont
  {Vasuth}}]{Gergely:1998sr}%
  \BibitemOpen
  \bibfield  {author} {\bibinfo {author} {\bibfnamefont {L.~A.}\ \bibnamefont
  {Gergely}}, \bibinfo {author} {\bibfnamefont {Z.~I.}\ \bibnamefont {Perjes}},
  \ and\ \bibinfo {author} {\bibfnamefont {M.}~\bibnamefont {Vasuth}},\ }\href
  {\doibase 10.1103/PhysRevD.58.124001} {\bibfield  {journal} {\bibinfo
  {journal} {Phys. Rev. D}\ }\textbf {\bibinfo {volume} {58}},\ \bibinfo
  {pages} {124001} (\bibinfo {year} {1998})},\ \Eprint
  {http://arxiv.org/abs/gr-qc/9808063} {arXiv:gr-qc/9808063} \BibitemShut
  {NoStop}%
\bibitem [{\citenamefont {Klein}\ and\ \citenamefont
  {Jetzer}(2010)}]{Klein:2010ti}%
  \BibitemOpen
  \bibfield  {author} {\bibinfo {author} {\bibfnamefont {A.}~\bibnamefont
  {Klein}}\ and\ \bibinfo {author} {\bibfnamefont {P.}~\bibnamefont {Jetzer}},\
  }\href {\doibase 10.1103/PhysRevD.81.124001} {\bibfield  {journal} {\bibinfo
  {journal} {Phys. Rev. D}\ }\textbf {\bibinfo {volume} {81}},\ \bibinfo
  {pages} {124001} (\bibinfo {year} {2010})},\ \Eprint
  {http://arxiv.org/abs/1005.2046} {arXiv:1005.2046 [gr-qc]} \BibitemShut
  {NoStop}%
\bibitem [{\citenamefont {Klein}\ \emph {et~al.}(2018)\citenamefont {Klein},
  \citenamefont {Boetzel}, \citenamefont {Gopakumar}, \citenamefont {Jetzer},\
  and\ \citenamefont {de~Vittori}}]{Klein:2018ybm}%
  \BibitemOpen
  \bibfield  {author} {\bibinfo {author} {\bibfnamefont {A.}~\bibnamefont
  {Klein}}, \bibinfo {author} {\bibfnamefont {Y.}~\bibnamefont {Boetzel}},
  \bibinfo {author} {\bibfnamefont {A.}~\bibnamefont {Gopakumar}}, \bibinfo
  {author} {\bibfnamefont {P.}~\bibnamefont {Jetzer}}, \ and\ \bibinfo {author}
  {\bibfnamefont {L.}~\bibnamefont {de~Vittori}},\ }\href {\doibase
  10.1103/PhysRevD.98.104043} {\bibfield  {journal} {\bibinfo  {journal} {Phys.
  Rev. D}\ }\textbf {\bibinfo {volume} {98}},\ \bibinfo {pages} {104043}
  (\bibinfo {year} {2018})},\ \Eprint {http://arxiv.org/abs/1801.08542}
  {arXiv:1801.08542 [gr-qc]} \BibitemShut {NoStop}%
\bibitem [{\citenamefont {Phukon}\ \emph {et~al.}(2019)\citenamefont {Phukon},
  \citenamefont {Gupta}, \citenamefont {Bose},\ and\ \citenamefont
  {Jain}}]{Phukon:2019gfh}%
  \BibitemOpen
  \bibfield  {author} {\bibinfo {author} {\bibfnamefont {K.~S.}\ \bibnamefont
  {Phukon}}, \bibinfo {author} {\bibfnamefont {A.}~\bibnamefont {Gupta}},
  \bibinfo {author} {\bibfnamefont {S.}~\bibnamefont {Bose}}, \ and\ \bibinfo
  {author} {\bibfnamefont {P.}~\bibnamefont {Jain}},\ }\href {\doibase
  10.1103/PhysRevD.100.124008} {\bibfield  {journal} {\bibinfo  {journal}
  {Phys. Rev. D}\ }\textbf {\bibinfo {volume} {100}},\ \bibinfo {pages}
  {124008} (\bibinfo {year} {2019})},\ \Eprint
  {http://arxiv.org/abs/1904.03985} {arXiv:1904.03985 [gr-qc]} \BibitemShut
  {NoStop}%
\bibitem [{\citenamefont {Barausse}\ \emph {et~al.}(2016)\citenamefont
  {Barausse}, \citenamefont {Yunes},\ and\ \citenamefont
  {Chamberlain}}]{Barausse:2016eii}%
  \BibitemOpen
  \bibfield  {author} {\bibinfo {author} {\bibfnamefont {E.}~\bibnamefont
  {Barausse}}, \bibinfo {author} {\bibfnamefont {N.}~\bibnamefont {Yunes}}, \
  and\ \bibinfo {author} {\bibfnamefont {K.}~\bibnamefont {Chamberlain}},\
  }\href {\doibase 10.1103/PhysRevLett.116.241104} {\bibfield  {journal}
  {\bibinfo  {journal} {Phys. Rev. Lett.}\ }\textbf {\bibinfo {volume} {116}},\
  \bibinfo {pages} {241104} (\bibinfo {year} {2016})},\ \Eprint
  {http://arxiv.org/abs/1603.04075} {arXiv:1603.04075 [gr-qc]} \BibitemShut
  {NoStop}%
\bibitem [{\citenamefont {Cardoso}\ \emph {et~al.}(2016)\citenamefont
  {Cardoso}, \citenamefont {Macedo}, \citenamefont {Pani},\ and\ \citenamefont
  {Ferrari}}]{Cardoso:2016olt}%
  \BibitemOpen
  \bibfield  {author} {\bibinfo {author} {\bibfnamefont {V.}~\bibnamefont
  {Cardoso}}, \bibinfo {author} {\bibfnamefont {C.~F.~B.}\ \bibnamefont
  {Macedo}}, \bibinfo {author} {\bibfnamefont {P.}~\bibnamefont {Pani}}, \ and\
  \bibinfo {author} {\bibfnamefont {V.}~\bibnamefont {Ferrari}},\ }\href
  {\doibase 10.1088/1475-7516/2016/05/054} {\bibfield  {journal} {\bibinfo
  {journal} {JCAP}\ }\textbf {\bibinfo {volume} {05}},\ \bibinfo {pages} {054}
  (\bibinfo {year} {2016})},\ \bibinfo {note} {[Erratum: JCAP 04, E01
  (2020)]},\ \Eprint {http://arxiv.org/abs/1604.07845} {arXiv:1604.07845
  [hep-ph]} \BibitemShut {NoStop}%
\bibitem [{\citenamefont {Brito}\ \emph {et~al.}(2015)\citenamefont {Brito},
  \citenamefont {Cardoso},\ and\ \citenamefont {Pani}}]{Brito:2015oca}%
  \BibitemOpen
  \bibfield  {author} {\bibinfo {author} {\bibfnamefont {R.}~\bibnamefont
  {Brito}}, \bibinfo {author} {\bibfnamefont {V.}~\bibnamefont {Cardoso}}, \
  and\ \bibinfo {author} {\bibfnamefont {P.}~\bibnamefont {Pani}},\ }\href
  {\doibase 10.1007/978-3-319-19000-6} {\emph {\bibinfo {title}
  {{Superradiance}: {Energy Extraction, Black-Hole Bombs and Implications for
  Astrophysics and Particle Physics}}}},\ Vol.\ \bibinfo {volume} {906}\
  (\bibinfo  {publisher} {Springer},\ \bibinfo {year} {2015})\ \Eprint
  {http://arxiv.org/abs/1501.06570} {arXiv:1501.06570 [gr-qc]} \BibitemShut
  {NoStop}%
\bibitem [{\citenamefont {Cardoso}\ \emph {et~al.}(2011)\citenamefont
  {Cardoso}, \citenamefont {Chakrabarti}, \citenamefont {Pani}, \citenamefont
  {Berti},\ and\ \citenamefont {Gualtieri}}]{Cardoso:2011xi}%
  \BibitemOpen
  \bibfield  {author} {\bibinfo {author} {\bibfnamefont {V.}~\bibnamefont
  {Cardoso}}, \bibinfo {author} {\bibfnamefont {S.}~\bibnamefont
  {Chakrabarti}}, \bibinfo {author} {\bibfnamefont {P.}~\bibnamefont {Pani}},
  \bibinfo {author} {\bibfnamefont {E.}~\bibnamefont {Berti}}, \ and\ \bibinfo
  {author} {\bibfnamefont {L.}~\bibnamefont {Gualtieri}},\ }\href {\doibase
  10.1103/PhysRevLett.107.241101} {\bibfield  {journal} {\bibinfo  {journal}
  {Phys.\ Rev.\ Lett.}\ }\textbf {\bibinfo {volume} {107}},\ \bibinfo {pages}
  {241101} (\bibinfo {year} {2011})},\ \Eprint {http://arxiv.org/abs/1109.6021}
  {arXiv:1109.6021 [gr-qc]} \BibitemShut {NoStop}%
\bibitem [{\citenamefont {Yunes}\ \emph {et~al.}(2012)\citenamefont {Yunes},
  \citenamefont {Pani},\ and\ \citenamefont {Cardoso}}]{Yunes:2011aa}%
  \BibitemOpen
  \bibfield  {author} {\bibinfo {author} {\bibfnamefont {N.}~\bibnamefont
  {Yunes}}, \bibinfo {author} {\bibfnamefont {P.}~\bibnamefont {Pani}}, \ and\
  \bibinfo {author} {\bibfnamefont {V.}~\bibnamefont {Cardoso}},\ }\href
  {\doibase 10.1103/PhysRevD.85.102003} {\bibfield  {journal} {\bibinfo
  {journal} {Phys. Rev. D}\ }\textbf {\bibinfo {volume} {85}},\ \bibinfo
  {pages} {102003} (\bibinfo {year} {2012})},\ \Eprint
  {http://arxiv.org/abs/1112.3351} {arXiv:1112.3351 [gr-qc]} \BibitemShut
  {NoStop}%
\bibitem [{\citenamefont {Cardoso}\ \emph {et~al.}(2019)\citenamefont
  {Cardoso}, \citenamefont {del Rio},\ and\ \citenamefont
  {Kimura}}]{Cardoso:2019nis}%
  \BibitemOpen
  \bibfield  {author} {\bibinfo {author} {\bibfnamefont {V.}~\bibnamefont
  {Cardoso}}, \bibinfo {author} {\bibfnamefont {A.}~\bibnamefont {del Rio}}, \
  and\ \bibinfo {author} {\bibfnamefont {M.}~\bibnamefont {Kimura}},\
  }\href@noop {} {\  (\bibinfo {year} {2019})},\ \Eprint
  {http://arxiv.org/abs/1907.01561} {arXiv:1907.01561 [gr-qc]} \BibitemShut
  {NoStop}%
%%CITATION = ARXIV:1907.01561;%%
\bibitem [{\citenamefont {Christiansen}\ \emph {et~al.}(2020)\citenamefont
  {Christiansen}, \citenamefont {Jim\'enez},\ and\ \citenamefont
  {Mota}}]{Christiansen:2020pnv}%
  \BibitemOpen
  \bibfield  {author} {\bibinfo {author} {\bibfnamefont {O.}~\bibnamefont
  {Christiansen}}, \bibinfo {author} {\bibfnamefont {J.~B.}\ \bibnamefont
  {Jim\'enez}}, \ and\ \bibinfo {author} {\bibfnamefont {D.~F.}\ \bibnamefont
  {Mota}},\ }\href@noop {} {\  (\bibinfo {year} {2020})},\ \Eprint
  {http://arxiv.org/abs/2003.11452} {arXiv:2003.11452 [gr-qc]} \BibitemShut
  {NoStop}%
\bibitem [{\citenamefont {Liu}\ \emph {et~al.}(2020)\citenamefont {Liu},
  \citenamefont {Guo}, \citenamefont {Cai},\ and\ \citenamefont
  {Kim}}]{Liu:2020cds}%
  \BibitemOpen
  \bibfield  {author} {\bibinfo {author} {\bibfnamefont {L.}~\bibnamefont
  {Liu}}, \bibinfo {author} {\bibfnamefont {Z.-K.}\ \bibnamefont {Guo}},
  \bibinfo {author} {\bibfnamefont {R.-G.}\ \bibnamefont {Cai}}, \ and\
  \bibinfo {author} {\bibfnamefont {S.~P.}\ \bibnamefont {Kim}},\ }\href
  {\doibase 10.1103/PhysRevD.102.043508} {\bibfield  {journal} {\bibinfo
  {journal} {Phys. Rev. D}\ }\textbf {\bibinfo {volume} {102}},\ \bibinfo
  {pages} {043508} (\bibinfo {year} {2020})},\ \Eprint
  {http://arxiv.org/abs/2001.02984} {arXiv:2001.02984 [astro-ph.CO]}
  \BibitemShut {NoStop}%
\bibitem [{\citenamefont {Cardoso}\ \emph {et~al.}(2020)\citenamefont
  {Cardoso}, \citenamefont {Guo}, \citenamefont {Macedo},\ and\ \citenamefont
  {Pani}}]{Cardoso:2020nst}%
  \BibitemOpen
  \bibfield  {author} {\bibinfo {author} {\bibfnamefont {V.}~\bibnamefont
  {Cardoso}}, \bibinfo {author} {\bibfnamefont {W.-d.}\ \bibnamefont {Guo}},
  \bibinfo {author} {\bibfnamefont {C.~F.}\ \bibnamefont {Macedo}}, \ and\
  \bibinfo {author} {\bibfnamefont {P.}~\bibnamefont {Pani}},\ }\href@noop {}
  {\  (\bibinfo {year} {2020})},\ \Eprint {http://arxiv.org/abs/2009.07287}
  {arXiv:2009.07287 [gr-qc]} \BibitemShut {NoStop}%
\bibitem [{\citenamefont {Bondi}\ and\ \citenamefont
  {Hoyle}(1944)}]{Bondi:1944jm}%
  \BibitemOpen
  \bibfield  {author} {\bibinfo {author} {\bibfnamefont {H.}~\bibnamefont
  {Bondi}}\ and\ \bibinfo {author} {\bibfnamefont {F.}~\bibnamefont {Hoyle}},\
  }\href@noop {} {\bibfield  {journal} {\bibinfo  {journal} {Mon. Not. Roy.
  Astron. Soc.}\ }\textbf {\bibinfo {volume} {104}},\ \bibinfo {pages} {273}
  (\bibinfo {year} {1944})}\BibitemShut {NoStop}%
\bibitem [{\citenamefont {Macedo}\ \emph {et~al.}(2013)\citenamefont {Macedo},
  \citenamefont {Pani}, \citenamefont {Cardoso},\ and\ \citenamefont
  {Crispino}}]{Macedo:2013qea}%
  \BibitemOpen
  \bibfield  {author} {\bibinfo {author} {\bibfnamefont {C.~F.~B.}\
  \bibnamefont {Macedo}}, \bibinfo {author} {\bibfnamefont {P.}~\bibnamefont
  {Pani}}, \bibinfo {author} {\bibfnamefont {V.}~\bibnamefont {Cardoso}}, \
  and\ \bibinfo {author} {\bibfnamefont {L.~C.~B.}\ \bibnamefont {Crispino}},\
  }\href {\doibase 10.1088/0004-637X/774/1/48} {\bibfield  {journal} {\bibinfo
  {journal} {Astrophys. J.}\ }\textbf {\bibinfo {volume} {774}},\ \bibinfo
  {pages} {48} (\bibinfo {year} {2013})},\ \Eprint
  {http://arxiv.org/abs/1302.2646} {arXiv:1302.2646 [gr-qc]} \BibitemShut
  {NoStop}%
%%CITATION = ARXIV:1302.2646;%%
\bibitem [{\citenamefont {Edgar}(2004)}]{Edgar:2004mk}%
  \BibitemOpen
  \bibfield  {author} {\bibinfo {author} {\bibfnamefont {R.~G.}\ \bibnamefont
  {Edgar}},\ }\href {\doibase 10.1016/j.newar.2004.06.001} {\bibfield
  {journal} {\bibinfo  {journal} {New Astron. Rev.}\ }\textbf {\bibinfo
  {volume} {48}},\ \bibinfo {pages} {843} (\bibinfo {year} {2004})},\ \Eprint
  {http://arxiv.org/abs/astro-ph/0406166} {arXiv:astro-ph/0406166} \BibitemShut
  {NoStop}%
\bibitem [{\citenamefont {Chandrasekhar}(1943)}]{Chandrasekhar:1943ys}%
  \BibitemOpen
  \bibfield  {author} {\bibinfo {author} {\bibfnamefont {S.}~\bibnamefont
  {Chandrasekhar}},\ }\href {\doibase 10.1086/144517} {\bibfield  {journal}
  {\bibinfo  {journal} {Astrophys. J.}\ }\textbf {\bibinfo {volume} {97}},\
  \bibinfo {pages} {255} (\bibinfo {year} {1943})}\BibitemShut {NoStop}%
\bibitem [{\citenamefont {Ostriker}(1999)}]{Ostriker:1998fa}%
  \BibitemOpen
  \bibfield  {author} {\bibinfo {author} {\bibfnamefont {E.~C.}\ \bibnamefont
  {Ostriker}},\ }\href {\doibase 10.1086/306858} {\bibfield  {journal}
  {\bibinfo  {journal} {Astrophys. J.}\ }\textbf {\bibinfo {volume} {513}},\
  \bibinfo {pages} {252} (\bibinfo {year} {1999})},\ \Eprint
  {http://arxiv.org/abs/astro-ph/9810324} {arXiv:astro-ph/9810324 [astro-ph]}
  \BibitemShut {NoStop}%
%%CITATION = ASTRO-PH/9810324;%%
\bibitem [{\citenamefont {Annulli}\ \emph {et~al.}(2020)\citenamefont
  {Annulli}, \citenamefont {Cardoso},\ and\ \citenamefont
  {Vicente}}]{Annulli:2020lyc}%
  \BibitemOpen
  \bibfield  {author} {\bibinfo {author} {\bibfnamefont {L.}~\bibnamefont
  {Annulli}}, \bibinfo {author} {\bibfnamefont {V.}~\bibnamefont {Cardoso}}, \
  and\ \bibinfo {author} {\bibfnamefont {R.}~\bibnamefont {Vicente}},\ }\href
  {\doibase 10.1103/PhysRevD.102.063022} {\bibfield  {journal} {\bibinfo
  {journal} {Phys. Rev. D}\ }\textbf {\bibinfo {volume} {102}},\ \bibinfo
  {pages} {063022} (\bibinfo {year} {2020})},\ \Eprint
  {http://arxiv.org/abs/2009.00012} {arXiv:2009.00012 [gr-qc]} \BibitemShut
  {NoStop}%
\bibitem [{\citenamefont {{Dokuchaev}}(1964)}]{Dokuchaev:1964}%
  \BibitemOpen
  \bibfield  {author} {\bibinfo {author} {\bibfnamefont {V.~P.}\ \bibnamefont
  {{Dokuchaev}}},\ }\href@noop {} {\bibfield  {journal} {\bibinfo  {journal}
  {\sovast}\ }\textbf {\bibinfo {volume} {8}},\ \bibinfo {pages} {23} (\bibinfo
  {year} {1964})}\BibitemShut {NoStop}%
\bibitem [{\citenamefont {{Ruderman}}\ and\ \citenamefont
  {{Spiegel}}(1971)}]{Ruderman:1971}%
  \BibitemOpen
  \bibfield  {author} {\bibinfo {author} {\bibfnamefont {M.~A.}\ \bibnamefont
  {{Ruderman}}}\ and\ \bibinfo {author} {\bibfnamefont {E.~A.}\ \bibnamefont
  {{Spiegel}}},\ }\href {\doibase 10.1086/150870} {\bibfield  {journal}
  {\bibinfo  {journal} {\apj}\ }\textbf {\bibinfo {volume} {165}},\ \bibinfo
  {pages} {1} (\bibinfo {year} {1971})}\BibitemShut {NoStop}%
\bibitem [{\citenamefont {{Rephaeli}}\ and\ \citenamefont
  {{Salpeter}}(1980)}]{Rephaeli:1980}%
  \BibitemOpen
  \bibfield  {author} {\bibinfo {author} {\bibfnamefont {Y.}~\bibnamefont
  {{Rephaeli}}}\ and\ \bibinfo {author} {\bibfnamefont {E.~E.}\ \bibnamefont
  {{Salpeter}}},\ }\href {\doibase 10.1086/158202} {\bibfield  {journal}
  {\bibinfo  {journal} {\apj}\ }\textbf {\bibinfo {volume} {240}},\ \bibinfo
  {pages} {20} (\bibinfo {year} {1980})}\BibitemShut {NoStop}%
\bibitem [{\citenamefont {Antoni}\ \emph {et~al.}(2019)\citenamefont {Antoni},
  \citenamefont {MacLeod},\ and\ \citenamefont
  {Ramirez-Ruiz}}]{Antoni:2019pgq}%
  \BibitemOpen
  \bibfield  {author} {\bibinfo {author} {\bibfnamefont {A.}~\bibnamefont
  {Antoni}}, \bibinfo {author} {\bibfnamefont {M.}~\bibnamefont {MacLeod}}, \
  and\ \bibinfo {author} {\bibfnamefont {E.}~\bibnamefont {Ramirez-Ruiz}},\
  }\href {\doibase 10.3847/1538-4357/ab3466} {\bibfield  {journal} {\bibinfo
  {journal} {Astrophys. J.}\ }\textbf {\bibinfo {volume} {884}},\ \bibinfo
  {pages} {22} (\bibinfo {year} {2019})},\ \Eprint
  {http://arxiv.org/abs/1901.07572} {arXiv:1901.07572 [astro-ph.HE]}
  \BibitemShut {NoStop}%
\bibitem [{\citenamefont {Vicente}\ \emph {et~al.}(2019)\citenamefont
  {Vicente}, \citenamefont {Cardoso},\ and\ \citenamefont
  {Zilhao}}]{Vicente:2019ilr}%
  \BibitemOpen
  \bibfield  {author} {\bibinfo {author} {\bibfnamefont {R.}~\bibnamefont
  {Vicente}}, \bibinfo {author} {\bibfnamefont {V.}~\bibnamefont {Cardoso}}, \
  and\ \bibinfo {author} {\bibfnamefont {M.}~\bibnamefont {Zilhao}},\ }\href
  {\doibase 10.1093/mnras/stz2526} {\bibfield  {journal} {\bibinfo  {journal}
  {Mon.\ Not.\ Roy.\ Astron.\ Soc.}\ }\textbf {\bibinfo {volume} {489}},\
  \bibinfo {pages} {5424} (\bibinfo {year} {2019})},\ \Eprint
  {http://arxiv.org/abs/1905.06353} {arXiv:1905.06353 [astro-ph.GA]}
  \BibitemShut {NoStop}%
\bibitem [{\citenamefont {Gair}\ \emph {et~al.}(2011)\citenamefont {Gair},
  \citenamefont {Flanagan}, \citenamefont {Drasco}, \citenamefont {Hinderer},\
  and\ \citenamefont {Babak}}]{Gair:2010iv}%
  \BibitemOpen
  \bibfield  {author} {\bibinfo {author} {\bibfnamefont {J.~R.}\ \bibnamefont
  {Gair}}, \bibinfo {author} {\bibfnamefont {E.~E.}\ \bibnamefont {Flanagan}},
  \bibinfo {author} {\bibfnamefont {S.}~\bibnamefont {Drasco}}, \bibinfo
  {author} {\bibfnamefont {T.}~\bibnamefont {Hinderer}}, \ and\ \bibinfo
  {author} {\bibfnamefont {S.}~\bibnamefont {Babak}},\ }\href {\doibase
  10.1103/PhysRevD.83.044037} {\bibfield  {journal} {\bibinfo  {journal} {Phys.
  Rev. D}\ }\textbf {\bibinfo {volume} {83}},\ \bibinfo {pages} {044037}
  (\bibinfo {year} {2011})},\ \Eprint {http://arxiv.org/abs/1012.5111}
  {arXiv:1012.5111 [gr-qc]} \BibitemShut {NoStop}%
\bibitem [{\citenamefont {Landau}\ and\ \citenamefont
  {Lifshitz}(1982)}]{landau1982mechanics}%
  \BibitemOpen
  \bibfield  {author} {\bibinfo {author} {\bibfnamefont {L.}~\bibnamefont
  {Landau}}\ and\ \bibinfo {author} {\bibfnamefont {E.}~\bibnamefont
  {Lifshitz}},\ }\href {https://books.google.pt/books?id=bE-9tUH2J2wC} {\emph
  {\bibinfo {title} {Mechanics}}},\ \bibinfo {number} {v. 1}\ (\bibinfo
  {publisher} {Elsevier Science},\ \bibinfo {year} {1982})\BibitemShut
  {NoStop}%
\bibitem [{\citenamefont {De~Luca}\ \emph {et~al.}(2020)\citenamefont
  {De~Luca}, \citenamefont {Franciolini}, \citenamefont {Pani},\ and\
  \citenamefont {Riotto}}]{DeLuca:2020qqa}%
  \BibitemOpen
  \bibfield  {author} {\bibinfo {author} {\bibfnamefont {V.}~\bibnamefont
  {De~Luca}}, \bibinfo {author} {\bibfnamefont {G.}~\bibnamefont
  {Franciolini}}, \bibinfo {author} {\bibfnamefont {P.}~\bibnamefont {Pani}}, \
  and\ \bibinfo {author} {\bibfnamefont {A.}~\bibnamefont {Riotto}},\ }\href
  {\doibase 10.1088/1475-7516/2020/06/044} {\bibfield  {journal} {\bibinfo
  {journal} {JCAP}\ }\textbf {\bibinfo {volume} {06}},\ \bibinfo {pages} {044}
  (\bibinfo {year} {2020})},\ \Eprint {http://arxiv.org/abs/2005.05641}
  {arXiv:2005.05641 [astro-ph.CO]} \BibitemShut {NoStop}%
\bibitem [{\citenamefont {{Salmassi}}(1985)}]{1985Salmassi}%
  \BibitemOpen
  \bibfield  {author} {\bibinfo {author} {\bibfnamefont {M.}~\bibnamefont
  {{Salmassi}}},\ }\href {\doibase 10.1007/BF01261625} {\bibfield  {journal}
  {\bibinfo  {journal} {Celestial Mechanics}\ }\textbf {\bibinfo {volume}
  {37}},\ \bibinfo {pages} {359} (\bibinfo {year} {1985})}\BibitemShut
  {NoStop}%
\bibitem [{\citenamefont {{Djukic}}(1993)}]{1993Djukic}%
  \BibitemOpen
  \bibfield  {author} {\bibinfo {author} {\bibfnamefont {D.~S.}\ \bibnamefont
  {{Djukic}}},\ }\href {\doibase 10.1007/BF00696184} {\bibfield  {journal}
  {\bibinfo  {journal} {Celestial Mechanics and Dynamical Astronomy}\ }\textbf
  {\bibinfo {volume} {56}},\ \bibinfo {pages} {523} (\bibinfo {year}
  {1993})}\BibitemShut {NoStop}%
\bibitem [{\citenamefont {Amaro-Seoane}\ \emph {et~al.}(2017)\citenamefont
  {Amaro-Seoane} \emph {et~al.}}]{Audley:2017drz}%
  \BibitemOpen
  \bibfield  {author} {\bibinfo {author} {\bibfnamefont {P.}~\bibnamefont
  {Amaro-Seoane}} \emph {et~al.} (\bibinfo {collaboration} {LISA}),\
  }\href@noop {} {\  (\bibinfo {year} {2017})},\ \Eprint
  {http://arxiv.org/abs/1702.00786} {arXiv:1702.00786 [astro-ph.IM]}
  \BibitemShut {NoStop}%
\bibitem [{\citenamefont {Maggiore}(2008)}]{MaggioreBook}%
  \BibitemOpen
  \bibfield  {author} {\bibinfo {author} {\bibfnamefont {M.}~\bibnamefont
  {Maggiore}},\ }\href@noop {} {\emph {\bibinfo {title} {Gravitational Waves:
  Volume 1: Theory and Experiments}}}\ (\bibinfo  {publisher} {Oxford
  University Press},\ \bibinfo {address} {Oxford},\ \bibinfo {year}
  {2008})\BibitemShut {NoStop}%
\bibitem [{\citenamefont {Hopper}\ and\ \citenamefont
  {Cardoso}(2018)}]{Hopper:2017qus}%
  \BibitemOpen
  \bibfield  {author} {\bibinfo {author} {\bibfnamefont {S.}~\bibnamefont
  {Hopper}}\ and\ \bibinfo {author} {\bibfnamefont {V.}~\bibnamefont
  {Cardoso}},\ }\href {\doibase 10.1103/PhysRevD.97.044031} {\bibfield
  {journal} {\bibinfo  {journal} {Phys. Rev. D}\ }\textbf {\bibinfo {volume}
  {97}},\ \bibinfo {pages} {044031} (\bibinfo {year} {2018})},\ \Eprint
  {http://arxiv.org/abs/1706.02791} {arXiv:1706.02791 [gr-qc]} \BibitemShut
  {NoStop}%
\bibitem [{\citenamefont {Nitz}\ and\ \citenamefont
  {Capano}(2020)}]{Nitz:2020mga}%
  \BibitemOpen
  \bibfield  {author} {\bibinfo {author} {\bibfnamefont {A.~H.}\ \bibnamefont
  {Nitz}}\ and\ \bibinfo {author} {\bibfnamefont {C.~D.}\ \bibnamefont
  {Capano}},\ }\href@noop {} {\  (\bibinfo {year} {2020})},\ \Eprint
  {http://arxiv.org/abs/2010.12558} {arXiv:2010.12558 [astro-ph.HE]}
  \BibitemShut {NoStop}%
\bibitem [{\citenamefont {Barausse}\ \emph {et~al.}(2014)\citenamefont
  {Barausse}, \citenamefont {Cardoso},\ and\ \citenamefont
  {Pani}}]{Barausse:2014tra}%
  \BibitemOpen
  \bibfield  {author} {\bibinfo {author} {\bibfnamefont {E.}~\bibnamefont
  {Barausse}}, \bibinfo {author} {\bibfnamefont {V.}~\bibnamefont {Cardoso}}, \
  and\ \bibinfo {author} {\bibfnamefont {P.}~\bibnamefont {Pani}},\ }\href
  {\doibase 10.1103/PhysRevD.89.104059} {\bibfield  {journal} {\bibinfo
  {journal} {Phys. Rev.}\ }\textbf {\bibinfo {volume} {D89}},\ \bibinfo {pages}
  {104059} (\bibinfo {year} {2014})},\ \Eprint {http://arxiv.org/abs/1404.7149}
  {arXiv:1404.7149 [gr-qc]} \BibitemShut {NoStop}%
%%CITATION = ARXIV:1404.7149;%%
\bibitem [{\citenamefont {Cardoso}\ and\ \citenamefont
  {Maselli}(2019)}]{Cardoso:2019rou}%
  \BibitemOpen
  \bibfield  {author} {\bibinfo {author} {\bibfnamefont {V.}~\bibnamefont
  {Cardoso}}\ and\ \bibinfo {author} {\bibfnamefont {A.}~\bibnamefont
  {Maselli}},\ }\href@noop {} {\  (\bibinfo {year} {2019})},\ \Eprint
  {http://arxiv.org/abs/1909.05870} {arXiv:1909.05870 [astro-ph.HE]}
  \BibitemShut {NoStop}%
\bibitem [{\citenamefont {Toubiana}\ \emph {et~al.}(2020)\citenamefont
  {Toubiana} \emph {et~al.}}]{Toubiana:2020drf}%
  \BibitemOpen
  \bibfield  {author} {\bibinfo {author} {\bibfnamefont {A.}~\bibnamefont
  {Toubiana}} \emph {et~al.},\ }\href@noop {} {\  (\bibinfo {year} {2020})},\
  \Eprint {http://arxiv.org/abs/2010.06056} {arXiv:2010.06056 [astro-ph.HE]}
  \BibitemShut {NoStop}%
\bibitem [{\citenamefont {Cardoso}\ and\ \citenamefont
  {Duque}(2020)}]{Cardoso:2019upw}%
  \BibitemOpen
  \bibfield  {author} {\bibinfo {author} {\bibfnamefont {V.}~\bibnamefont
  {Cardoso}}\ and\ \bibinfo {author} {\bibfnamefont {F.}~\bibnamefont
  {Duque}},\ }\href {\doibase 10.1103/PhysRevD.101.064028} {\bibfield
  {journal} {\bibinfo  {journal} {Phys. Rev. D}\ }\textbf {\bibinfo {volume}
  {101}},\ \bibinfo {pages} {064028} (\bibinfo {year} {2020})},\ \Eprint
  {http://arxiv.org/abs/1912.07616} {arXiv:1912.07616 [gr-qc]} \BibitemShut
  {NoStop}%
\bibitem [{\citenamefont {Roedig}\ and\ \citenamefont
  {Sesana}(2012)}]{Roedig:2011rn}%
  \BibitemOpen
  \bibfield  {author} {\bibinfo {author} {\bibfnamefont {C.}~\bibnamefont
  {Roedig}}\ and\ \bibinfo {author} {\bibfnamefont {A.}~\bibnamefont
  {Sesana}},\ }\href {\doibase 10.1088/1742-6596/363/1/012035} {\bibfield
  {journal} {\bibinfo  {journal} {J. Phys. Conf. Ser.}\ }\textbf {\bibinfo
  {volume} {363}},\ \bibinfo {pages} {012035} (\bibinfo {year} {2012})},\
  \Eprint {http://arxiv.org/abs/1111.3742} {arXiv:1111.3742 [astro-ph.CO]}
  \BibitemShut {NoStop}%
\bibitem [{\citenamefont {Zrake}\ \emph {et~al.}(2020)\citenamefont {Zrake},
  \citenamefont {Tiede}, \citenamefont {MacFadyen},\ and\ \citenamefont
  {Haiman}}]{Zrake:2020zkw}%
  \BibitemOpen
  \bibfield  {author} {\bibinfo {author} {\bibfnamefont {J.}~\bibnamefont
  {Zrake}}, \bibinfo {author} {\bibfnamefont {C.}~\bibnamefont {Tiede}},
  \bibinfo {author} {\bibfnamefont {A.}~\bibnamefont {MacFadyen}}, \ and\
  \bibinfo {author} {\bibfnamefont {Z.}~\bibnamefont {Haiman}},\ }\href@noop {}
  {\  (\bibinfo {year} {2020})},\ \Eprint {http://arxiv.org/abs/2010.09707}
  {arXiv:2010.09707 [astro-ph.HE]} \BibitemShut {NoStop}%
\bibitem [{\citenamefont {Burko}\ \emph {et~al.}(2002)\citenamefont {Burko},
  \citenamefont {Harte},\ and\ \citenamefont {Poisson}}]{Burko:2002ge}%
  \BibitemOpen
  \bibfield  {author} {\bibinfo {author} {\bibfnamefont {L.~M.}\ \bibnamefont
  {Burko}}, \bibinfo {author} {\bibfnamefont {A.~I.}\ \bibnamefont {Harte}}, \
  and\ \bibinfo {author} {\bibfnamefont {E.}~\bibnamefont {Poisson}},\ }\href
  {\doibase 10.1103/PhysRevD.65.124006} {\bibfield  {journal} {\bibinfo
  {journal} {Phys. Rev. D}\ }\textbf {\bibinfo {volume} {65}},\ \bibinfo
  {pages} {124006} (\bibinfo {year} {2002})},\ \Eprint
  {http://arxiv.org/abs/gr-qc/0201020} {arXiv:gr-qc/0201020} \BibitemShut
  {NoStop}%
\bibitem [{\citenamefont {Quinn}(2000)}]{Quinn:2000wa}%
  \BibitemOpen
  \bibfield  {author} {\bibinfo {author} {\bibfnamefont {T.~C.}\ \bibnamefont
  {Quinn}},\ }\href {\doibase 10.1103/PhysRevD.62.064029} {\bibfield  {journal}
  {\bibinfo  {journal} {Phys. Rev. D}\ }\textbf {\bibinfo {volume} {62}},\
  \bibinfo {pages} {064029} (\bibinfo {year} {2000})},\ \Eprint
  {http://arxiv.org/abs/gr-qc/0005030} {arXiv:gr-qc/0005030} \BibitemShut
  {NoStop}%
\bibitem [{\citenamefont {Poisson}\ and\ \citenamefont
  {Will}(2014)}]{Poisson_will_2014}%
  \BibitemOpen
  \bibfield  {author} {\bibinfo {author} {\bibfnamefont {E.}~\bibnamefont
  {Poisson}}\ and\ \bibinfo {author} {\bibfnamefont {C.~M.}\ \bibnamefont
  {Will}},\ }\href {\doibase 10.1017/CBO9781139507486} {\emph {\bibinfo {title}
  {Gravity: Newtonian, Post-Newtonian, Relativistic}}}\ (\bibinfo  {publisher}
  {Cambridge University Press},\ \bibinfo {year} {2014})\BibitemShut {NoStop}%
\end{thebibliography}%
\end{document}